\documentclass[aps, prd, reprint, twocolumn, nofootinbib, superscriptaddress]{revtex4-2}
\usepackage{amsmath}
\usepackage{amssymb}
\usepackage{amsfonts}
\usepackage{amsthm}
\usepackage{mathrsfs}
\usepackage{booktabs, array}
\usepackage{soul}
\usepackage[T1]{fontenc}
\usepackage[caption=false]{subfig}
\usepackage{graphicx} 
\usepackage{multirow}
\usepackage{makecell}
\usepackage[dvipsnames]{xcolor}
\usepackage{tikz}

\newcommand{\geo}{{\mbox{\tiny geo}}}

\definecolor{linkcolor}{rgb}{0.0,0.3,0.5}
\usepackage[
    hypertexnames=false,
    unicode,
    colorlinks=true,
    linkcolor=linkcolor,
    citecolor=linkcolor,
    filecolor=linkcolor,
    urlcolor=linkcolor,
    pdfusetitle
]{hyperref}
\usepackage{url}
\hypersetup{breaklinks=true}

\usepackage{orcidlink}
\newcommand{\dCS}{{\mbox{\tiny dCS}}}

\newcommand{\UIUC}{Illinois  Center  for  Advanced  Studies  of  the  Universe \&
Department of Physics, University of Illinois at Urbana-Champaign, Urbana, Illinois 61801, USA}
\newcommand{\CAL}{Theoretical Astrophysics 350-17, California Institute of Technology, Pasadena, CA 91125, USA}

\allowdisplaybreaks
\begin{document}
    \title{Extreme mass-ratio inspiral within an ultralight scalar cloud \\ I. Scalar radiation}
    \author{Dongjun Li}
    \affiliation{\UIUC}
    \affiliation{\CAL}

    \author{Colin Weller}
    \affiliation{\CAL}
    
    \author{Patrick Bourg}
    \affiliation{Institute for Mathematics, Astrophysics and Particle Physics, Radboud University, Heyendaalseweg 135, 6525 AJ Nijmegen, The Netherlands}
    
    \author{Michael LaHaye}
    \affiliation{Department of Physics, University of Guelph, Guelph, Ontario, N1G 2W1, Canada}

    \author{Nicol\'as Yunes}
    \affiliation{\UIUC}
    
    \author{Huan Yang}
    \email{hyangdoa@tsinghua.edu.cn} 
    \affiliation{Department of Astronomy, Tsinghua University, Beijing 100084, China}

    \date{\today}
    \begin{abstract}
        In this work, we study the dynamics of an extreme mass-ratio inspiral (EMRI) embedded within a scalar cloud populated around the massive black hole. This cloud may be generated through the black hole superradiant process if the wavelength of the scalar particle is comparable to the size of the massive black hole. The EMRI motion perturbs the cloud, producing scalar radiation towards infinity and into the black hole horizon. In addition, the backreaction of the scalar radiation onto the orbit modifies the motion of the EMRI and induces an observable gravitational-wave phase shift for a range of system parameters. We quantify the scalar flux and the induced phase shift, as one of the examples of exactly-solvable, environmental effects of EMRIs. 
        \end{abstract}
    \maketitle

\section{Introduction}

Since the first detection of gravitational waves (GWs) emitted by a binary black hole (BH) merger in 2015 \cite{LIGOScientific:2016aoc}, more than one hundred GW events from binary mergers have been detected \cite{KAGRA:2021vkt}. These detections offer novel opportunities to test Einstein's general relativity (GR) in the strong gravity regime \cite{Yunes:2016jcc, LIGOScientific:2016lio, Nair:2019iur, Silva:2020acr, Perkins:2021mhb, LIGOScientific:2021sio, Schumacher:2023cxh, Lagos:2024boe, Liu:2024atc} and probe the astrophysical environments these compact objects live in \cite{Barausse:2014pra, Barausse:2014tra, Cole:2022yzw}. Upcoming space-based GW detectors, such as LISA \cite{LISA:2017pwj, LISA:2024hlh}, Taiji \cite{Hu:2017mde, Ruan:2018tsw}, and TianQin \cite{TianQin:2015yph, TianQin:2020hid}, will expand the observable parameter space of BH mergers, including supermassive mergers and extreme mass-ratio inspirals (EMRIs) \cite{LISA:2024hlh}. The latter are composed of a central supermassive BH and an accompanying, secondary, stellar-mass compact object that zooms and whirls around the supermassive BH, emitting GWs in the process. 

Thanks to the hundreds of thousands of cycles contained in the GWs emitted during EMRIs \cite{Barack:2018yly}, one can map precisely the spacetime geometry of the central supermassive BH \cite{Ryan:1995wh, Ryan:1997hg, Barack:2006pq, Hinderer:2008dm, Vigeland:2009pr, Vigeland:2011ji, Fransen:2022jtw}, uncover possible signals of beyond-GR theories \cite{Sopuerta:2009iy, Yunes:2011aa, Pani:2011xj, Canizares:2012is, Stein:2013wza, Zimmerman:2015hua, Cardoso:2018zhm, Chua:2018yng, LISA:2022kgy, Tan:2024utr}, and examine the properties of the surrounding environments, such as accretion disks (i.e. wet EMRIs \cite{Pan:2021ksp,Pan:2021oob}) \cite{Barausse:2007dy, Yunes:2011ws, Kocsis:2011dr, Derdzinski:2018qzv, Speri:2022upm, Brito:2023pyl}, ultralight bosonic clouds \cite{Arvanitaki:2009fg, Arvanitaki:2010sy, Zhang:2018kib, Berti:2019wnn, Zhang:2019eid, Maselli:2020zgv, Traykova:2021dua, Baumann:2021fkf, Vicente:2022ivh, Baumann:2022pkl, Khalvati:2024tzz}, dark matter spikes or halos \cite{Hernquist:1990be, Gondolo:1999ef, Ullio:2001fb, Zhao:2005zr, Eda:2013gg, Eda:2014kra, Yue:2017iwc, Yue:2018vtk, Hannuksela:2019vip, Kavanagh:2020cfn, Cardoso:2022whc, Destounis:2022obl, Duque:2023seg, Gliorio:2025cbh, Mitra:2025tag}, or tidal resonance due to other nearby compact objects \cite{Yang:2017aht, Bonga:2019ycj, Yang:2019iqa}. To disentangle these effects and extract fundamental physics or astrophysics from precise EMRI measurements, accurate waveforms are necessary. The phase of these waves should be accurate to within one radian \cite{Hughes:2016xwf, LISAConsortiumWaveformWorkingGroup:2023arg, LISA:2024hlh}, when accounting for the additional effects discussed above.

\begin{figure}[t]
    \centering
    \includegraphics[width=\linewidth]{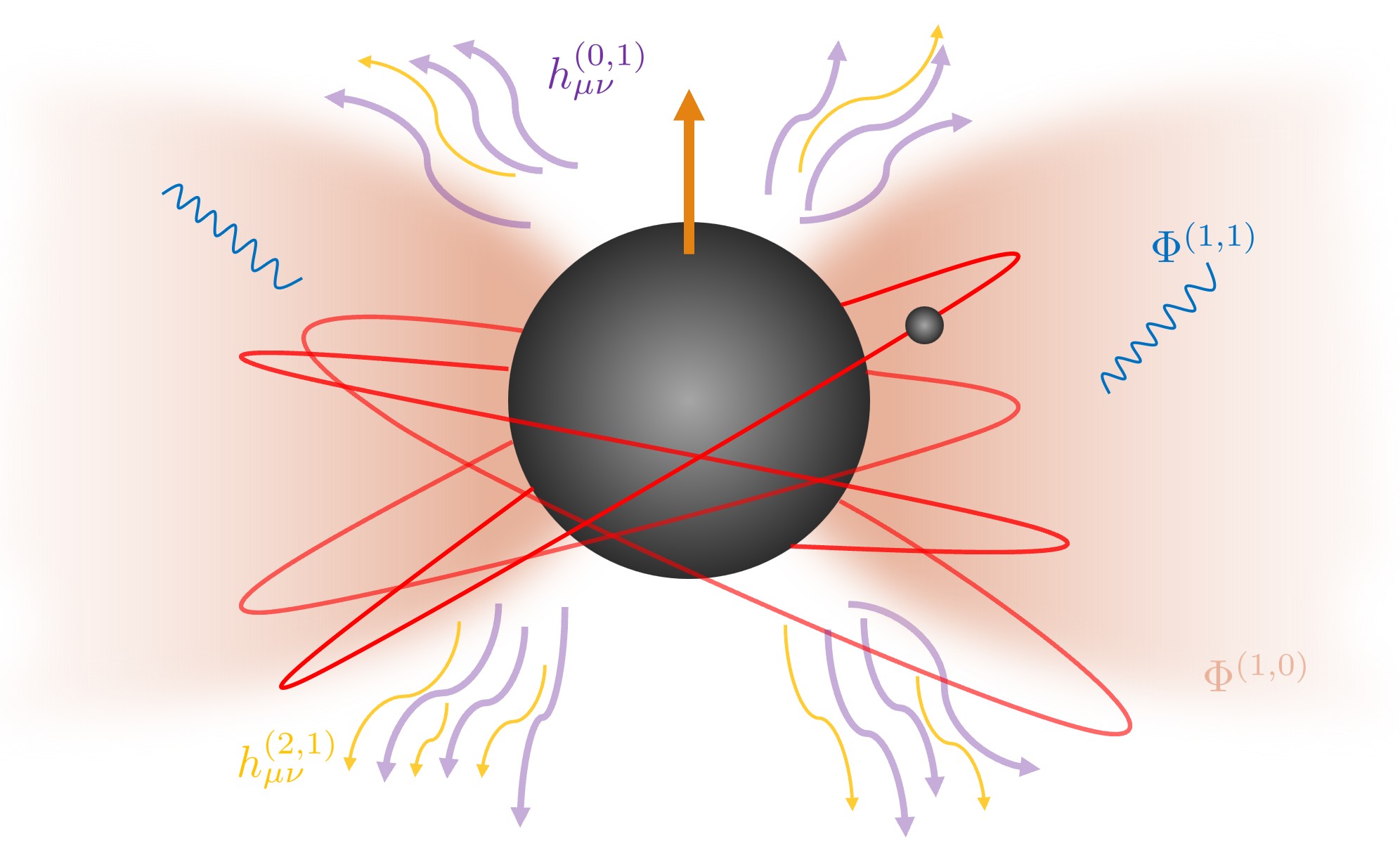}
    \caption{A schematic illustration of the system studied in this work. An EMRI, composed of a rotating supermassive BH (large black ball) and a secondary stellar-mass compact object (small black ball), is embedded in a complex ultralight scalar cloud formed via superradiance. The labels $\Phi^{(0,1)}$ and $\Phi^{(1,1)}$ denote the background scalar profile (pink blobs) and its radiation (blue curly lines), respectively. The labels $h_{\mu\nu}^{(0,1)}$ and $h_{\mu\nu}^{(2,1)}$ denote the GWs driven by the secondary in the absence of a cloud (purple arrows) and the additional GWs sourced by the scalar cloud (yellow arrows), respectively. The thick, orange arrow represents the spin angular momentum of the supermassive BH, while the red curve shows the trajectory of the secondary object.}
    \label{fig:EMRI_in_scalar_cloud}
\end{figure}

The state-of-the-art approach to generating waveforms for EMRIs uses self-force theory~\cite{Mino:1996nk, Hughes:1999bq, Lousto:2002em, Hinderer:2008dm, Hughes:2016xwf, Barack:2018yvs, Pound:2021qin}. Leveraging the fact that the mass ratio $\epsilon$ of the central BH to the secondary is a small parameter, one expands the field equations in powers of $\epsilon$. The field equations are then solved iteratively, order by order, where solutions of the lower-order equations are used as input into the source terms appearing in the higher-order equations. Much progress has been made in GR along these lines, where, at linear order in $\epsilon$, generic orbits of rotating BHs have been well-studied \cite{vandeMeent:2017bcc, Hughes:2021exa, Isoyama:2021jjd}, and efficient frameworks to model the generation of waveforms are also possible \cite{Chua:2017ujo, Chua:2020stf}. However, in order to achieve phase accuracy within one radian, second-order contributions also need to be incorporated \cite{Hughes:2016xwf, LISA:2022kgy, LISAConsortiumWaveformWorkingGroup:2023arg}. Due to the substantially increased difficulty involved in second-order self-force, most efforts have been restricted to a Schwarzschild background~\cite{Barack:2005nr, Barack:2007tm, Wardell:2021fyy, Bourg:2024vre, PanossoMacedo:2024pox}, with current calculations allowing for a spinning secondary (and a slowly spinning primary)~\cite{Pound:2019lzj, Warburton:2021kwk, Wardell:2021fyy, Mathews:2021rod}. Recently, much progress has been made in extending the calculations to Kerr~\cite{Dolan:2021ijg, Toomani:2021jlo, Dolan:2023enf, Wardell:2024yoi, Mathews:2025nyb}.

Due to radiation-reaction effects, the secondary slowly loses energy and angular momentum during the inspiral via the emission of GWs. To compute energy and angular momentum fluxes within such a system, the most popular approach is the one developed by Teukolsky in \cite{Teukolsky:1973ha, Press:1973zz, Teukolsky:1974yv}. Using the Newman-Penrose (NP) formalism \cite{Newman:1961qr}, this approach focuses on curvature perturbations caused by the small BH on the supermassive BH background. The perturbed curvature equations can be decoupled to find separable, second-order evolution equations for the Weyl scalars $\Psi_0$ and $\Psi_4$. From the solution to these \textit{Teukolsky equations} for $\Psi_0$ or $\Psi_4$, one can then directly compute the energy and angular momentum carried away by GWs into the horizon and out to spatial infinity, respectively.

\begin{figure}[t]
    \centering
    \includegraphics[width=\linewidth]{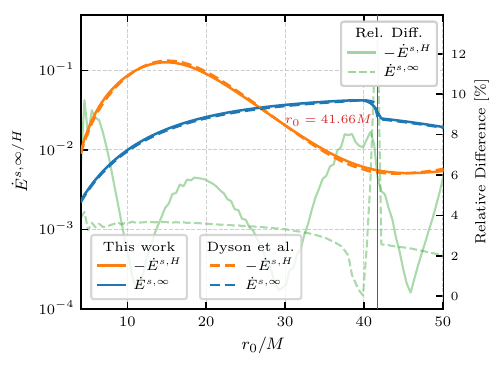}
    \caption{The infinity and horizon energy fluxes $\dot{E}^{s,\infty/H}$ (left axis) associated with the scalar radiation $\Phi^{(1,1)}$ of the $(\ell_c=m_c=1,n_c=0)$ complex scalar cloud around a Kerr BH of spin $a=0.88M$, where the complex scalar field has mass $\mu M=0.3$. The blue lines represent the infinity flux $\dot{E}^{s,\infty}$, while the orange lines show the horizon flux $\dot{E}^{s,H}$ but with an overall minus sign, given that $\dot{E}^{s,H}$ is always negative here. The solid and dashed lines are the fluxes obtained via the approach in this work and the one in \cite{Dyson:2025dlj}, respectively. The green solid and dashed lines show the relative fractional differences in percentages (right axis) between these two approaches in the horizon and infinity fluxes, respectively. The red vertical line marks the threshold radius $r_0$ where the infinity flux has a sharp decrease.}
    \label{fig:total_flux_11}
\end{figure}

The derivation of the Teukolsky equations requires the background BH spacetime to be Ricci-flat and of Petrov type D under Petrov classification \cite{Petrov:2000bs}, conditions satisfied by rotating BHs in GR \cite{Teukolsky:1973ha}. However, in beyond-GR theories or when in the presence of environmental effects, these conditions may no longer hold \cite{Yunes:2009hc, Yagi:2012ya, Herdeiro:2014goa, Kleihaus:2015aje, Bamber:2021knr, DeLuca:2021ite, Owen:2021eez, Fernandes:2022gde, Yagi:2023eap}, which has long hindered extending the Teukolsky formalism to such scenarios. The recently developed \textit{modified Teukolsky formalism} (MTF) \cite{Li:2022pcy, Hussain:2022ins} has the promise of overcoming this challenge. This formalism extends the Teukolsky formalism to non-Ricci-flat and algebraically-general BH backgrounds by making a two-parameter expansion of the NP equations, under which a set of Teukolsky-like (decoupled and separable) equations for $\Psi_0$ and $\Psi_4$ can be derived. The MTF has already seen some success in the study of ringdown in a few beyond-GR theories, such as higher-derivative gravity \cite{Cano:2023tmv, Cano:2023jbk, Cano:2024ezp, Maenaut:2024oci} and dynamical Chern-Simons (dCS) gravity \cite{Wagle:2023fwl, Li:2025fci}. Despite having been applied only to study BH ringdown in beyond-GR theories so far, the MTF can be used generically to investigate how ringdown and EMRIs are affected when the background BH spacetime is deformed perturbatively under beyond-GR corrections, astrophysical environments, or other effects. 

In this work, we take the first steps toward applying the MTF to study EMRIs in the presence of a specific environmental effect: an ultralight complex scalar cloud, as depicted in Fig.~\ref{fig:EMRI_in_scalar_cloud}. The studies of EMRIs embedded in ultralight scalar clouds have so far focused mostly on non-rotating supermassive BH backgrounds \cite{Duque:2023seg, Brito:2023pyl}, the post-Newtonian limit \cite{Baumann:2018vus, Berti:2019wnn, Zhang:2018kib, Zhang:2019eid}, or linear motions \cite{Traykova:2021dua, Vicente:2022ivh, Dyson:2024qrq}. Important progress has been recently made in \cite{Dyson:2025dlj} to model the scalar radiation from a scalar cloud perturbed by an EMRI around a rotating supermassive BH in a fully relativistic framework. Our goal is to advance this line of research through a series of works that develop a comprehensive MTF-based framework for computing both scalar and gravitational radiation from such systems, enabling GW phase evolution, full waveform generation, and the exploration of their observational signatures.

As a first step, we extend the MTF to EMRIs around ``dirty'' BHs in astrophysical environments. Using this extension, we compute the scalar radiation emitted by a complex scalar cloud perturbed by a stellar-mass object in a circular and equatorial orbit around a rotating supermassive BH, evaluate the associated energy and angular momentum fluxes, and compare our results with those of \cite{Dyson:2025dlj}. Our semi-analytical approach, though independent from the more numerical methods in \cite{Dyson:2025dlj}, yields consistent results for the scalar radiation profile and fluxes at the horizon and infinity (i.e., see Fig.~\ref{fig:total_flux_11} for a comparison of the fluxes) and is more efficient given the analytical simplifications made.\footnote{We have also corrected two minor errors in \cite{Dyson:2025dlj}: the spheroidal separation constant of the scalar field equation and the normalization factor for the horizon flux.} In addition to the dipolar clouds studied in \cite{Dyson:2025dlj}, we also extend this analysis to quadrupolar clouds and show that different cloud configurations can affect the secondary's inspiral differently. For example, the quadrupolar cloud's scalar radiation accelerates the secondary's inspiral at small orbital radii, whereas the dipolar cloud slows it down. This work establishes key techniques for our forthcoming study on gravitational radiation and provides essential scalar data for constructing the source term in the modified Teukolsky equation.

The remainder of this work presents the above calculations in detail. In Sec.~\ref{sec:EMRI_scalar_cloud}, we build the framework for studying EMRIs within an ultralight scalar cloud based on the MTF and classify the source terms in both the scalar- and gravitational-sector equations. In Sec.~\ref{sec:evaluation_scalar_eq}, we apply the Lorenz-gauge reconstruction approach in \cite{Dolan:2021ijg, Dolan:2023enf, Wardell:2024yoi} to construct the source term of the scalar radiation equation and extract its radial part, which we solve in Sec.~\ref{sec:scalar_radiation} using Green's function for dipolar and quadrupolar clouds around a supermassive Kerr BH. In Sec.~\ref{sec:fluxes}, we compute the energy and angular momentum fluxes from the scalar radiation profile in Sec.~\ref{sec:scalar_radiation}, and we discuss the future directions of this work in Sec.~\ref{sec:discussion}. Throughout this work, we adopt the following conventions unless stated otherwise: we work in 4-dimensions with metric signature $(-,+,+,+)$ as in \cite{Misner:1973prb}. For all NP quantities except the metric signature, we use the notation adapted by Chandrasekhar in \cite{Chandrasekhar_1983}. We set $M=1$ when plotting our results.

\section{EMRI within an ultralight scalar cloud}
\label{sec:EMRI_scalar_cloud}

A complex scalar field $\Phi$ with mass $\mu$ and minimally coupled to gravity in GR is described by the action \cite{Brito:2023pyl}
\begin{equation}
    S=\int d^4 x \sqrt{-g}\left(\frac{R}{16\pi}
    -\partial^\mu\Phi\partial_\mu\bar{\Phi}
    -\mu^2\Phi\bar{\Phi}+\mathcal{L}_{\mathrm{m}}\right)\,,
\end{equation}
where $\mathcal{L}_{\mathrm{m}}$ is the Lagrangian density of additional, non-minimally coupled matter fields, and an overhead bar stands for complex conjugation. The corresponding equations of motion are given by
\begin{align}
    \square\Phi &= \mu^2\Phi\,, \label{eq:EOM_scalar} \\
    G_{\mu\nu}
    & =8\pi\left(T_{\mu\nu}^{\Phi}+T^{m}_{\mu\nu}\right)\,, 
    \label{eq:EOM_h}
\end{align}
where the stress-energy tensor of the complex scalar field $\Phi$ is 
\begin{equation} \label{eq:stress_Phi}
    T_{\mu\nu}^{\Phi}
    =2\partial_{(\mu}\Phi\partial_{\nu)}\bar{\Phi}-g_{\mu\nu}
    \left(\partial_\alpha\Phi\partial^\alpha\bar{\Phi}
    +\mu^2\Phi\bar{\Phi}\right)\,, 
\end{equation}
and $T^{m}_{\mu\nu}$ is the stress-energy tensor of any additional matter. 

The main goal of this work is to study how an EMRI system evolves when it is embedded in a complex scalar field cloud. In the absence of a companion, the spacetime is assumed to contain a rotating (Kerr) BH of mass $M$ (i.e.~the additional matter stress-energy tensor is zero) and a quasi-bound state of a complex scalar field (or scalar cloud for short) of mass $M_c$ that grows through superradiance until reaching a stationary configuration. When considering a companion that enters this cloud, we set $T^{m}_{\mu\nu}=T^{p}_{\mu\nu}$, where the latter is the stress-energy tensor of the secondary object in the EMRI system, 
\begin{equation} \label{eq:stress_particle}
    T_{\mu\nu}^{p}
    =m_p\int u_{\mu}u_{\nu}\frac{\delta^{(4)}
    \left(x^\mu-x_p^\mu(\tau)\right)}{\sqrt{-g}}d\tau\,.
\end{equation}
In this equation, $m_p$ is the mass of the secondary object, $u^\mu \equiv dx_p^\mu/d\tau$ is its four-velocity normalized via $u_\mu u^\mu=-1$ for a time-like trajectory, $x_p^\mu(\tau)$ is its worldline, and $\tau$ is proper time.

Let us now follow the framework in \cite{Li:2022pcy, Hussain:2022ins, Cano:2023tmv, Cano:2023jbk, Brito:2023pyl, Dyson:2025dlj} and introduce two expansion parameters: the small mass-ratio $\epsilon=m_p/M$, that parametrizes how strongly the supermassive BH is perturbed by the secondary, and $\zeta$, that characterizes the amplitude of the complex scalar field\footnote{Reference \cite{Brito:2023pyl} has used $q$ for the small mass ratio and $\epsilon$ for the complex scalar field amplitude.}. Using these two parameters, the complex scalar field $\Phi$ and the metric $g_{\mu\nu}$ can be expanded as follows:
\begin{subequations} \label{eq:expansion_scheme}
\begin{align}
    \Phi 
    =& \;\left(\zeta\Phi^{(1,0)}+\mathcal{O}(\zeta^3)\right)
    +\epsilon\left(\zeta\Phi^{(1,1)}+\mathcal{O}(\zeta^3)\right)
    +\mathcal{O}(\epsilon^2)\,, \label{eq:expansion_scalar}\\
    g_{\mu\nu}
    =& \;\left(g_{\mu\nu}^{(0,0)}+\zeta^2 h_{\mu\nu}^{(2,0)}+\mathcal{O}(\zeta^4)\right) \nonumber\\
    & \;+\epsilon\left(h_{\mu\nu}^{(0,1)}+\zeta^2h_{\mu\nu}^{(2,1)}
    +\mathcal{O}(\zeta^4)\right)+\mathcal{O}(\epsilon^2)\,.
    \label{eq:expansion_metric}
\end{align}
\end{subequations}
In the expansion of $\Phi$ in Eq.~\eqref{eq:expansion_scalar}, $\Phi^{(1,0)}$ represents the ``background'' quasi-bound state of the scalar field, while $\Phi^{(1,1)}$ represents the scalar radiation, or the leading-order perturbation of the background scalar cloud, generated by the secondary compact object. In Eq.~\eqref{eq:expansion_metric}, $g_{\mu\nu}^{(0,0)}$ is the background BH metric in GR, which is the Kerr metric in our case. The term $h_{\mu\nu}^{(2,0)}$ is the deformation of the Kerr metric by the scalar cloud (i.e.~by $\Phi^{(1,0)}$). The term $h_{\mu\nu}^{(0,1)}$ is the gravitational radiation sourced by the secondary's linear perturbation of the supermassive BH, while $h_{\mu\nu}^{(2,1)}$ is the additional gravitational radiation due to the scalar cloud.

One key assumption of the expansion in Eq.~\eqref{eq:expansion_scalar} is that we only consider ultralight scalar clouds that perturbatively affect the central supermassive BH, e.g., its characteristic mass and density are much smaller than those of the supermassive BH. Thus, the expansion of $\Phi$ enters at $\mathcal{O}(\zeta^1)$. Following \cite{Dyson:2025dlj}, we choose $\zeta$ to be the ratio between the characteristic density of the scalar cloud and that of the Kerr BH such that $\zeta=(\mu M)^{3}\sqrt{M_c/M}$, where recall that $M$ is the mass of the supermassive BH and $M_c$ is the total mass of the scalar cloud. The latter is defined by \cite{Dyson:2025dlj}
\begin{equation} \label{eq:zeta_def}
    M_{c}=-\int_{r_{+}}^{\infty}\int_{S^2}
    g^{t\mu(0,0)}T_{\mu t}^{\Phi(1,0)}
    \left(r^2+a^2\cos^2{\theta}\right) dr\,d\Omega\,.
\end{equation}
The definition in Eq.~\eqref{eq:zeta_def} may look cyclic, but it is not: the stress-energy tensor of Eq.~\eqref{eq:stress_Phi} that enters Eq.~\eqref{eq:zeta_def} is to be evaluated using the Kerr metric $g_{\mu\nu}^{(0,0)}$ and the quasi-bound state of the scalar field $\Phi^{(1,0)}$ with a certain normalization, which then determines $\zeta$ for this normalization choice. 

Some previous work chose a different convention for the dimensionless constant $\zeta$ that characterizes the strength of beyond-GR or environmental effects. For example, in dCS gravity, Refs.~\cite{Yunes:2009hc, Yagi:2012ya, Wagle:2021tam, Srivastava:2021imr, Wagle:2023fwl, Li:2025fci} chose $\zeta\propto\alpha_{\dCS}^2$, where $\alpha_{\dCS}$ is the coupling constant for the dCS interaction term and the order at which the pseudoscalar field enters. In this work, we follow the conventions in \cite{Brito:2023pyl} by setting $\zeta$ at the order where the scalar field $\Phi$ enters. Since $T^{\Phi}_{\mu\nu}\sim\Phi^2$ according to Eq.~\eqref{eq:stress_Phi}, the corrections to the background BH spacetime start at $\mathcal{O}(\zeta^{2})$. Following arguments similar to those of other scalar-tensor theories \cite{Yunes:2009hc}, one can show that $\Phi$ is always driven by the metric perturbation at one order lower in $\zeta$, which, in turn, is used as input to compute the metric perturbation at one order higher in $\zeta$ \cite{Brito:2023pyl}. This hierarchical structure explains why perturbations of the scalar field all enter at odd powers of $\zeta$, while perturbations of the metric all appear at even powers.

Now, let us focus on the equations of motion in Eqs.~\eqref{eq:EOM_scalar} and \eqref{eq:EOM_h} and apply the expansion scheme in Eq.~\eqref{eq:expansion_scheme} to them. On the scalar side, one can expand Eq.~\eqref{eq:EOM_scalar} as \cite{Brito:2023pyl}
\begin{subequations} \label{eq:EOM_scalar_expanded}
\begin{align}
    & \left(\square^{(0,0)}-\mu^2\right)\Phi^{(1,0)}=0\,, 
    \label{eq:EOM_scalar_10} \\
    & \left(\square^{(0,0)}-\mu^2\right)\Phi^{(1,1)}
    =\mathcal{S}_{\Phi}^{(1,1)}\,, 
    \label{eq:EOM_scalar_11} \\
    & \mathcal{S}_{\Phi}^{(1,1)}
    =g^{\mu\nu(0,0)}\Gamma_{\mu \nu}^{\alpha(0,1)}\Phi_{;\alpha}^{(1,0)}
    +h^{\mu\nu(0,1)}\Phi_{;\mu\nu}^{(1,0)}\,,
    \label{eq:EOM_scalar_11_source}
\end{align}  
\end{subequations}
where $\Gamma^{\alpha}_{\mu\nu}$ are Christoffel symbols, and the semicolon in the above equations stands for covariant derivative with respect to the Kerr metric. The source term $\mathcal{S}_{\Phi}^{(1,1)}$ comes from linearizing the metric in the d'Alembertian operator $\square$ with respect to $\epsilon$, using Eq.~\eqref{eq:expansion_metric}. According to Eq.~\eqref{eq:EOM_scalar_11_source}, $\mathcal{S}_{\Phi}^{(1,1)}$ depends on the background scalar profile $\Phi^{(1,0)}$ and the metric perturbation $h_{\mu\nu}^{(0,1)}$ associated with the GWs driven by the secondary in GR. As shown in more detail in Sec.~\ref{sec:evaluation_scalar_eq}, $\Phi^{(1,0)}$ can be obtained via Leaver's method \cite{Leaver:1985ax, Dolan:2007mj}, while $h_{\mu\nu}^{(0,1)}$ can be calculated from the Lorenz-gauge reconstruction approach in \cite{Dolan:2021ijg, Dolan:2023enf, Wardell:2024yoi}. While Eq.~\eqref{eq:EOM_scalar_11} admits homogeneous solutions, we only consider the particular solutions driven by the source term $\mathcal{S}_{\Phi}^{(1,1)}$ in this work. This is because the quasi-bound states of the homogeneous solutions to Eq.~\eqref{eq:EOM_scalar_11} are already incorporated in the background scalar profile obtained from Eq.~\eqref{eq:EOM_scalar_10}, while the other quasinormal mode solutions are transient, given the timescale of the inspiral.

On the gravitational side, instead of solving the linearized Einstein equation from Eq.~\eqref{eq:EOM_h} directly, we choose to solve the associated modified Teukolsky equation in \cite{Li:2022pcy}, which is essentially a projection of Eq.~\eqref{eq:EOM_h} to the NP basis \cite{Hussain:2022ins}. The derivation of this equation can be found in \cite{Li:2022pcy}, where a short review of the NP formalism is also provided, while more comprehensive reviews of the NP formalism can be found in \cite{Newman:1961qr, Chandrasekhar_1983, Pound:2021qin}. Using the results in \cite{Li:2022pcy}, we find the modified Teukolsky equation of $\Psi_0$ in this particular case to be
\begin{subequations}
\begin{align}
    & H_{0}^{(0,0)}\Psi_{0}^{(0,1)}=\mathcal{S}_{p}^{(0,1)}\,, 
    \label{eq:Teuk_01} \\
    & H_{0}^{(0,0)}\Psi_{0}^{(2,1)}
    =\mathcal{S}_{\geo}^{(2,1)}+\mathcal{S}_{p}^{(2,1)}
    +\mathcal{S}_{T_{\Phi}}^{(2,1)}\,,
    \label{eq:Teuk_21}
\end{align}  
\end{subequations}
where $H_{0}$ is the Teukolsky operator, i.e.,
\begin{align} \label{eq:def_operators}
    & H_0=\mathcal{E}_2 F_2-\mathcal{E}_1F_1-3\Psi_2\,,\\
    & \mathcal{E}_1=E_1-\Psi_2^{-1}\delta\Psi_2\,,\quad
    \mathcal{E}_2=E_2-\Psi_2^{-1}D\Psi_2\,,  
\end{align}
with $\delta$ and $D$ being NP directional derivatives, while the operators $F_{1,2}$, $J_{1,2}$, and $E_{1,2}$ are defined as
\begin{equation} \label{eq:auxiliary_operators}
\begin{aligned}
    & F_1\equiv\bar{\delta}_{[-4,0,1,0]}\,,\quad
    && F_2\equiv\boldsymbol{\Delta}_{[1,0,-4,0]}\,, \\
    & J_1\equiv D_{[-2,0,-4,0]}\,,\quad
    && J_2\equiv\delta_{[0,-2,0,-4]}\,, \\
    & E_1\equiv\delta_{[-1,-3,1,-1]}\,,\quad
    && E_2\equiv D_{[-3,1,-1,-1]}\,.
\end{aligned}
\end{equation}
Here, we have used the shorthand notation  
\begin{align} \label{eq:operators_shorthand}
    D_{[a,b,c,d]}
    &=D+a\varepsilon+b\bar{\varepsilon}+c\rho+d\bar{\rho}\,, \nonumber\\
    \boldsymbol{\Delta}_{[a,b,c,d]}
    &=\boldsymbol{\Delta}+a\mu+b\bar{\mu}+c\gamma+d\bar{\gamma}\,, \nonumber\\
    \delta_{[a,b,c,d]}
    &=\delta+a\bar{\alpha}+b\beta+c\bar{\pi}+d\tau\,, \nonumber\\
    \bar{\delta}_{[a,b,c,d]}
    &=\bar{\delta}+a\alpha+b\bar{\beta}+c\pi+d\bar{\tau}\,,
\end{align}
where $\boldsymbol{\Delta}$ is another NP directional derivative, and $(\varepsilon,\rho,\mu,\gamma,\alpha,\beta,\pi,\tau)$ are (complex) spin coefficients. When evaluating the operators in Eq.~\eqref{eq:operators_shorthand} on the Kerr background, they can be directly mapped to the Chandrasekhar operators $\left\{\mathcal{D}_{n},\mathcal{D}^{\dagger}_{n},
\mathcal{L}_{n},\mathcal{L}^{\dagger}_{n}\right\}$ \cite{Chandrasekhar_1983, Dolan:2023enf, Ma:2024qcv}, as listed in Appendix~\ref{appendix:chandrasekhar_operators}.

At $\mathcal{O}(\zeta^0,\epsilon^1)$, the source term of the Teukolsky equation in Eq.~\eqref{eq:Teuk_01} is given by
\begin{equation}
    \mathcal{S}_{p}^{(0,1)}=\mathcal{E}_2^{(0,0)}S_{2,p}^{(0,1)}
    -\mathcal{E}_1^{(0,0)}S_{1,p}^{(0,1)}\,,
\end{equation}
where $S_{1,2}$ are sources in the Bianchi identities and determined by the NP Ricci scalars $\Phi_{ij}$ via
\begin{subequations} \label{eq:source_bianchi}
\begin{align}
    \label{eq:source_bianchi_1}
    \begin{split} 
        S_1\equiv& \;\delta_{[-2,-2,1,0]}\Phi_{00}
        -D_{[-2,0,0,-2]}\Phi_{01} \\
        & \;+2\sigma\Phi_{10}-2\kappa\Phi_{11}-\bar{\kappa}\Phi_{02}\,,
    \end{split} \\
    \label{eq:source_bianchi_2}
    \begin{split} 
        S_2\equiv& \;\delta_{[0,-2,2,0]}\Phi_{01}
        -D_{[-2,2,0,-1]}\Phi_{02} \\
        & \;-\bar{\lambda}\Phi_{00}+2\sigma\Phi_{11}-2\kappa\Phi_{12}\,.
    \end{split}
\end{align}
\end{subequations}
The relation between the stress-energy tensor and the NP Ricci scalars can be found in \cite{Chandrasekhar_1983, Li:2022pcy}. One can directly evaluate $\mathcal{S}_{p}^{(0,1)}$ using the stress-energy tensor $T_{\mu\nu}^{p}$ in Eq.~\eqref{eq:stress_particle}, and its explicit form for a circular orbit can be found in \cite{Hughes:1999bq}. The metric perturbation $h_{\mu\nu}^{(0,1)}$ can then be reconstructed from the Weyl scalars $\Psi_{0,4}^{(0,1)}$ via the Lorenz-gauge reconstruction method developed in \cite{Dolan:2021ijg, Dolan:2023enf, Wardell:2024yoi}. Since \cite{Dolan:2023enf} has only implemented this reconstruction procedure for circular and equatorial orbits, we also restrict attention to these orbits for the rest of this work.

At $\mathcal{O}(\zeta^2,\epsilon^1)$, the source terms of the modified Teukolsky equation in Eq.~\eqref{eq:Teuk_21} are given by
\begin{subequations} \label{eq:source_def_21}
\begin{align} 
    \mathcal{S}_{\geo}^{(2,1)}
    =& \;-H_0^{(2,0)}\Psi_0^{(0,1)}-H_0^{(0,1)}\Psi_0^{(2,0)}\,,
    \label{eq:S_geo} \\
    \mathcal{S}_{p}^{(2,1)}
    =& \;\mathcal{E}_2^{(0,0)}S_{2,p}^{(2,1)}+
    \mathcal{E}_2^{(2,0)}S_{2,p}^{(0,1)}
    -\mathcal{E}_1^{(0,0)}S_{1,p}^{(2,1)} \nonumber\\
    & \;-\mathcal{E}_1^{(2,0)}S_{1,p}^{(0,1)}\,, \label{eq:S_p}\\
    \mathcal{S}_{T_{\Phi}}^{(2,1)}
    =& \;\mathcal{E}_2^{(0,0)}S_{2,T_{\Phi}}^{(2,1)}+
    \mathcal{E}_2^{(0,1)}S_{2,T_{\Phi}}^{(2,0)}
    -\mathcal{E}_1^{(0,0)}S_{1,T_{\Phi}}^{(2,1)} \nonumber\\
    & \;-\mathcal{E}_1^{(0,1)}S_{1,T_{\Phi}}^{(2,0)}\,, \label{eq:S_bGR}
\end{align}
\end{subequations}
where $\mathcal{S}_{\geo}^{(2,1)}$ is driven by the correction to the background BH spacetime due to the scalar cloud, the latter of which has been studied in \cite{Herdeiro:2014goa} for nonlinear scalar clouds. The source terms $\mathcal{S}_{p}^{(2,1)}$ and $\mathcal{S}_{T_{\Phi}}^{(2,1)}$ are driven by the stress-energy tensor of the secondary $T_{\mu\nu}^{p}$ in Eq.~\eqref{eq:stress_particle} and the stress-energy tensor of the complex scalar field $T_{\mu\nu}^{\Phi}$ in Eq.~\eqref{eq:stress_Phi}, respectively. To evaluate $\mathcal{S}_{p}^{(2,1)}$ and $\mathcal{S}_{T_{\Phi}}^{(2,1)}$, one computes the corresponding $\Phi_{ij}$ from the stress-energy tensors $T_{\mu\nu}^{p}$ and $T_{\mu\nu}^{\Phi}$, respectively, inserts $\Phi_{ij}$ into Eqs.~\eqref{eq:source_bianchi} and \eqref{eq:source_def_21}, and linearizes. One feature worth noting in Eqs.~\eqref{eq:S_p} and \eqref{eq:S_bGR} is that $\mathcal{S}_{p}^{(2,1)}$ does not contain $S_{1,p}^{(2,0)}$ and $S_{2,p}^{(2,0)}$, while $\mathcal{S}_{T_{\Phi}}^{(2,1)}$ does not contain $S_{1,T_{\phi}}^{(0,1)}$  and $S_{2,T_{\phi}}^{(0,1)}$. This is because the stress-energy tensor of the secondary $T_{\mu\nu}^{p}$ starts at $\mathcal{O}(\epsilon^1)$, while the expansion of the scalar field in Eq.~\eqref{eq:expansion_scalar} starts at $\mathcal{O}(\zeta^1)$.

In this work, we will not solve the modified Teukolsky equation in Eq.~\eqref{eq:Teuk_21}\footnote{This calculation will be presented in detail in a follow-up work.} but focus on the scalar field equation in Eq.~\eqref{eq:EOM_scalar_11} instead. Here, we only provide a brief prescription of how to evaluate the sources in Eq.~\eqref{eq:Teuk_21}. From Eq.~\eqref{eq:source_def_21}, we notice that the source terms in the modified Teukolsky equation in Eq.~\eqref{eq:Teuk_21} take the form
\begin{subequations} \label{eq:coupling_struct}
\begin{align} 
    & \mathcal{S}_{\geo}^{(2,1)}
    \sim h_{\mu\nu}^{(2,0)}h_{\alpha\beta}^{(0,1)}\,, \label{eq:coupling_struct_S_geo} \\
    & \mathcal{S}_p^{(2,1)}
    \sim h_{\mu\nu}^{(2,0)}T_{\alpha\beta}^{p(0,1)}\,,
    \label{eq:coupling_struct_S_p} \\
    & \mathcal{S}_{T_{\Phi}}^{(2,1)}
    \sim h_{\mu\nu}^{(0,1)}\left(\Phi^{(1,0)}\right)^2
    +\Phi^{(1,0)}\Phi^{(1,1)}\,,
    \label{eq:coupling_struct_S_Tphi}
\end{align}  
\end{subequations}
where we have hidden any terms at $\mathcal{O}(\zeta^0,\epsilon^0)$. The term $T^{p(0,1)}_{\mu\nu}$ is the stress-energy tensor $T_{\mu\nu}^{p}$ evaluated on the GR background spacetime $g_{\mu\nu}^{(0,0)}$. Each field within Eq.~\eqref{eq:coupling_struct} can represent itself and its derivatives. For example, $h_{\mu\nu}^{(2,0)}h_{\alpha\beta}^{(0,1)}$ represents a coupling between $h_{\mu\nu}^{(2,0)}$, $h_{\mu\nu}^{(0,1)}$, and their derivatives. In principle, terms coupled to $T_{\mu\nu}^{p(2,1)}$ can also be obtained when evaluating $\mathcal{S}_p^{(2,1)}$, where $T_{\mu\nu}^{p(2,1)}$ is the first-order correction to $T^{p}_{\mu\nu}$ due to the background metric correction $h_{\mu\nu}^{(2,0)}$. However, we can expand $T_{\mu\nu}^{p(2,1)}$ in terms of $T_{\mu\nu}^{p(0,1)}$ after linearizing the background metric in terms of $g_{\mu\nu}^{(0,0)}$ and $h_{\mu\nu}^{(2,0)}$, so we get the same type of term in Eq.~\eqref{eq:coupling_struct_S_p}. 

From Eq.~\eqref{eq:coupling_struct}, we notice that the quantities we need for studying the gravitational sector are
\begin{equation}
    \Phi^{(1,0)}\,,\;\Phi^{(1,1)}\,,\;
    h_{\mu\nu}^{(2,0)}\,,\;h_{\mu\nu}^{(0,1)}\,.
\end{equation}
Although the source terms of the modified Teukolsky equation in Eq.~\eqref{eq:source_def_21} are much more complicated than the source term $S_{\Phi}^{(1,1)}$ of the scalar field equation in Eq.~\eqref{eq:EOM_scalar_11_source}, most of the infrastructure necessary for studying the gravitational sector is built up in this work for studying the scalar sector. As discussed above, this work aims to solve for $\Phi^{(1,1)}$, so we need to apply Leaver's method \cite{Leaver:1985ax, Dolan:2007mj} and Lorenz-gauge reconstruction \cite{Dolan:2021ijg, Dolan:2023enf, Wardell:2024yoi} to calculate $\Phi^{(1,0)}$ and $h_{\mu\nu}^{(0,1)}$, respectively, for evaluating $S_{\Phi}^{(1,1)}$. Besides developing a more effective strategy to evaluate the more complicated operators acting on $h_{\mu\nu}^{(0,1)}$ in Eq.~\eqref{eq:source_def_21}, the only additional information we need to evaluate the modified Teukolsky equation is the background metric correction $h_{\mu\nu}^{(2,0)}$. Fortunately, the non-linear backreaction of the cloud onto a Kerr geometry has been studied in \cite{Herdeiro:2014goa}, while additional efforts are required to incorporate the numerical background metric into the modified Teukolsky equation.

For the rest of this work, we focus on solving Eq.~\eqref{eq:EOM_scalar_11} to obtain the scalar radiation field $\Phi^{(1,1)}$ that is driven by the EMRI. We can then include the additional fluxes of energy and angular momentum carried away by $\Phi^{(1,1)}$ in the evolution of EMRIs within ultralight scalar clouds to study the impact these fluxes have on the GWs emitted. Furthermore, as shown in Eq.~\eqref{eq:coupling_struct}, we also need $\Phi^{(1,1)}$ to obtain the gravitational radiation $\Psi_{0,4}^{(1,1)}$ and its associated fluxes. Thus, this work represents a crucial first step towards completely modeling the GWs generated by an EMRI inside an ultralight scalar cloud.

\section{Evaluation of the scalar equation}
\label{sec:evaluation_scalar_eq}

In this section, we evaluate the scalar radiation equation [Eq.~\eqref{eq:EOM_scalar_11}] with the source in Eq.~\eqref{eq:EOM_scalar_11_source}. To calculate Eq.~\eqref{eq:EOM_scalar_11_source}, we need to know $\Phi^{(1,0)}$ and $h_{\mu\nu}^{(0,1)}$. The former represents quasi-bound states of the scalar cloud, which satisfy Eq.~\eqref{eq:EOM_scalar_10} and have been extensively studied in \cite{Dolan:2007mj}. Following \cite{Dolan:2007mj}, one can decompose $\Phi^{(1,0)}$ into harmonic modes, i.e.,
\begin{equation} \label{eq:phi_10_decomp}
    \Phi^{(1,0)}=\sum_{\ell_c,m_c}{}_{0}R_{\ell_c m_c\omega_c}^{\Phi}(r)\;
    {}_{0}S^{\Phi}_{\ell_c m_c\omega_c}(\theta)\;e^{-i\omega_c t+im_c\phi}\,,
\end{equation}
where we use the subscript $c$ to label the modes of the scalar cloud specifically and drop the additional subscript $n_c$ labeling the overtones for simplicity. The angular part ${}_{0}S^{\Phi}_{\ell_c m_c\omega_c}(\theta)$ and the radial part ${}_{0}R_{\ell_c m_c\omega_c}^{\Phi}(r)$ satisfy the angular and radial Teukolsky equations of a spin-$0$ massive particle, respectively. When doing so, one can solve for these quantities, as well as for $\omega_c$, via Leaver's method~\cite{Leaver:1985ax, Dolan:2007mj}. 

In this work, we will independently investigate the scalar radiation of the first two dominant superradiant modes forming quasi-bound clouds, so we will not sum over different $(\ell_c,m_c,n_c)$ modes. As we will see in Sec.~\ref{sec:fluxes}, a cloud with higher $\ell_c$ and $n_c$ generally leads to a peak in its profile that appears at a larger radius, so such clouds usually have subdominant effects on the secondary's evolution within the sensitivity band of space-based detectors, like LISA. One can also directly see this feature from the non-relativistic treatment of the cloud as a ``gravitational atom'' \cite{Dolan:2007mj, Baumann:2018vus, Berti:2019wnn}, where the Bohr radius of the cloud is given by 
\begin{equation} \label{eq:bohr_radius}
    r_{\mathrm{Bohr}}\approx\frac{M\left(\ell_c+n_c+1\right)^2}{(\mu M)^2}\,,
\end{equation}
which increases with both $\ell_c$ and $n_c$. 

\begin{figure}[t]
    \centering
    \includegraphics[width=\linewidth]{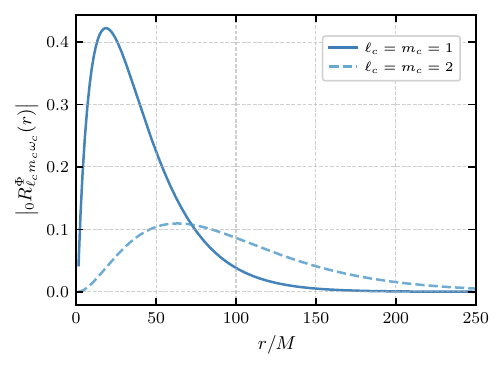}
    \caption{The radial part ${}_{0}R_{\ell_c m_c\omega_c}^{\Phi}(r)$ of the background scalar profile $\Phi^{(1,0)}$. The solid and dashed lines are for the fundamental modes of the dipolar ($\ell_c=m_c=1, n_c=0$) and quadrupolar clouds ($\ell_c=m_c=2, n_c=0$), respectively. Both clouds are around a rotating supermassive BH of spin $a=0.88M$ with the scalar mass $\mu M=0.3$.}
    \label{fig:scalar_profile}
\end{figure}

To compare our results fairly to those in \cite{Dyson:2025dlj}, we first consider the fundamental mode of a dipolar cloud ($\ell_c=m_c=1,n_c=0$) with scalar mass $\mu M=0.3$ around a Kerr BH of spin $a=0.88M$, such that its frequency $\omega_c$ satisfies
\begin{equation} \label{eq:omega_c_11}
    M\omega_c(\ell_c=m_c=1,n_c=0)\approx0.296294\,,
\end{equation}
which is obtained via Leaver's method in \cite{Dolan:2007mj}. Besides the dipolar cloud, we also study the next dominant superradiant mode, the fundamental mode of a quadrupolar cloud $(\ell_{c}=m_{c}=2,n_c=0)$, where
\begin{equation} \label{eq:omega_c_22}
    M\omega_c(\ell_c=m_c=2,n_c=0)\approx0.298451\,.
\end{equation}
One expects that after the dominant superradiant mode decays away due to GW radiation, the next dominant superradiant mode takes over \cite{Arvanitaki:2014wva}. For both states, the angular part ${}_{0}S^{\Phi}_{\ell_c m_c\omega_c}(\theta)$ of Eq.~\eqref{eq:phi_10_decomp} is the standard spin-weighted spheroidal harmonics for a spin-$0$ particle with the spheroidicity $\gamma$ determined by $a$, $\mu$, and $\omega_c$ via \cite{Dolan:2007mj}
\begin{equation}
    \gamma=a\sqrt{\omega_c^2-\mu^2}\,.
\end{equation}
We use the \texttt{SpinWeightedSpheroidalHarmonics} package in \cite{BHPToolkit, wardell_2024_11199019} to compute ${}_{0}S^{\Phi}_{\ell_c m_c\omega_c}(\theta)$. The radial part ${}_{0}R_{\ell_c m_c\omega_c}^{\Phi}(r)$ can be computed via Leaver's method in \cite{Leaver:1985ax, Dolan:2007mj}, and we include the first $150$ terms in the continuous fraction. In Fig.~\ref{fig:scalar_profile}, we plot $\left|{}_{0}R_{\ell_c m_c\omega_c}^{\Phi}(r)\right|$ of these two cloud configurations with the frequencies $\omega_c$ given in Eqs.~\eqref{eq:omega_c_11} and \eqref{eq:omega_c_22}, respectively. Note that EMRIs detectable by LISA are typically at orbital radii $r_0 \lesssim 20 M$, so the secondary is always inside the $\ell_c=m_c=1$ cloud for equatorial orbits.

\begin{figure*}[t]
    \centering
    \includegraphics[width=\linewidth]{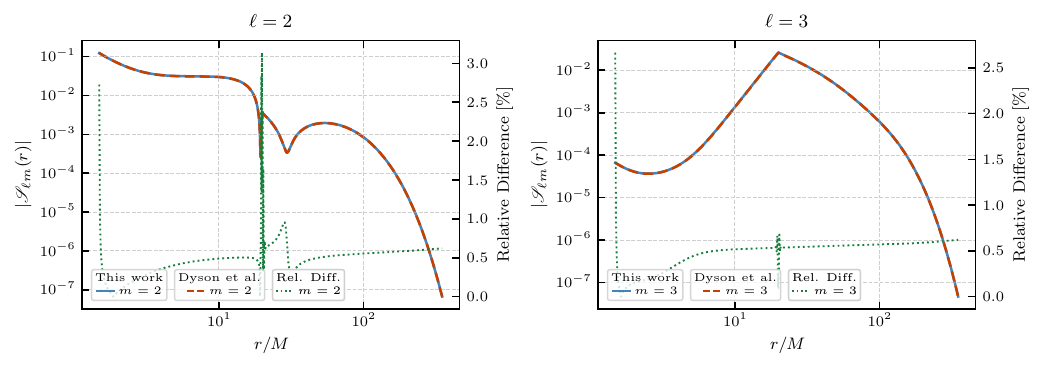}
    \caption{The radial source $\mathscr{S}_{\ell m}(r)$ (left axis) at $\ell=2,3$ and $r_0=20M$ for the $(\ell_c=m_c=1,n_c=0)$ scalar cloud around a Kerr BH of spin $a=0.88M$ and with scalar mass $\mu M=0.3$. As discussed in Sec.~\ref{sec:evaluation_scalar_eq}, the sources with $\ell$ and $m$ of opposite parity are zero, so they are not displayed here. We also only include positive $m$ for simplicity. The solid and dashed lines are the sources obtained via the approach in this work and the one in \cite{Dyson:2025dlj}, respectively. The dotted lines show the relative fractional differences in percentages (right axis) between these two approaches when computing $\left|\mathscr{S}_{\ell m}(r)\right|$.}
    \label{fig:source_l2l3}
\end{figure*}

To obtain $h_{\mu\nu}^{(0,1)}$, one needs to solve for $\Psi_{0,4}^{(0,1)}$ from Eq.~\eqref{eq:Teuk_01} and reconstruct the metric. For vacuum spacetime, a standard approach was developed by Chrzanowski, Cohen, Kegeles, and Ori \cite{Cohen_Kegeles_1975, Chrzanowski:1975wv, Kegeles:1979an, Ori:2002uv}, and it is known as the ``CCK-Ori procedure.'' Due to the radiation gauges chosen in this approach, one cannot directly apply it to non-vacuum spacetime, such as when an EMRI secondary is present. Nonetheless, much progress has been made to resolve this issue, such as the correction tensor approach developed in \cite{Green:2019nam, Toomani:2021jlo} or the Lorenz-gauge reconstruction developed in \cite{Dolan:2021ijg, Dolan:2023enf, Wardell:2024yoi}, both of which are essentially extensions of the CCK-Ori procedure to non-vacuum spacetimes. In this work, we use the Lorenz-gauge reconstruction developed in \cite{Dolan:2021ijg, Dolan:2023enf, Wardell:2024yoi}, as this approach has already been explicitly implemented for EMRIs with a secondary in the circular, equatorial orbit around a supermassive Kerr BH \cite{Dolan:2023enf}. In this case, $h_{\mu\nu}^{(0,1)}$ satisfies the Lorenz gauge condition
\begin{equation}
    \nabla^{\mu}\bar{h}_{\mu\nu}=0\,,
\end{equation}
where $\bar{h}_{\mu\nu}=h_{\mu\nu}-\frac{1}{2}g_{\mu\nu}h$, and $h$ is the trace of the metric perturbation with respect to the Kerr background. Since $g^{\mu\nu(0,0)}\Gamma_{\mu\nu}^{\alpha(0,1)}=\nabla^{\mu}\bar{h}_{\mu\alpha}$, the first term in Eq.~\eqref{eq:EOM_scalar_11_source} vanishes. 

When working with the modified Teukolsky equation and the reconstructed metric in \cite{Dolan:2023enf}, it is more convenient to work in the NP basis, where all geometric quantities are projected onto a tetrad that satisfies certain orthogonality conditions. For the case of Kerr BHs, a common choice of tetrad is the normalized Kinnersley tetrad $e_{a}^{\mu}=\{l^{\mu},n^{\mu},m^{\mu},\bar{m}^{\mu}\}$, where
\begin{align} \label{eq:normalized_tetrad}
    & l^\mu=l_{+}^\mu\,,\quad
    n^\mu=-\frac{\Delta(r)}{2\Gamma(r,\theta)
    \bar{\Gamma}(r,\theta)}l_{-}^\mu\,, \nonumber\\
    & m^\mu=\frac{1}{\sqrt{2}\Gamma(r,\theta)}m_{+}^\mu\,,\quad 
    \bar{m}^\mu=\frac{1}{\sqrt{2}\bar{\Gamma}(r,\theta)}m_{-}^\mu\,,
\end{align}
with $\tilde{e}_{a}^{\mu}=\{l_{+}^{\mu},l_{-}^{\mu},m_{+}^{\mu},m_{-}^{\mu}\}$ being the unnormalized Kinnersley tetrad used in \cite{Dolan:2023enf}, i.e.,
\begin{align} \label{eq:tetrad}
    l_{\pm}^\mu 
    & =\left[\pm\left(r^2+a^2\right)/\Delta(r),1,
    0,\pm a/\Delta(r)\right]\,, \nonumber\\
    m_{\pm}^\mu 
    & =[\pm ia\sin\theta,0,1,\pm i\csc\theta]\,,
\end{align}
where $\Delta(r)=r^2-2Mr+a^2$ and $\Gamma(r,\theta)=r+ia\cos{\theta}$. After projecting $\mathcal{S}_{\Phi}^{(1,1)}$ onto $e_{a}^{\mu}$, one finds
\begin{equation} \label{eq:EOM_scalar_11_source_Lorentz}
    \mathcal{S}_{\Phi}^{(1,1)}
    =h^{ab(0,1)}\Phi_{;ab}^{(1,0)}\,,
\end{equation}
where lowercase Latin indices stand for tetrad indices. In the following, those tetrad indices with a tilde are projected using the unnormalized tetrad $\tilde{e}_{a}^{\mu}$, while those without a tilde are projected using the normalized tetrad $e_{a}^{\mu}$. Following \cite{Chandrasekhar_1983}, we compute all the NP directional derivatives $\{D,\mathbf{\Delta},\delta,\bar{\delta}\}$ and spin coefficients using the normalized Kinnersley tetrad $e_{a}^{\mu}$, while the reconstructed metric $h_{\tilde{a}\tilde{b}}^{(0,1)}$ in \cite{Dolan:2023enf} is in the unnormalized Kinnersley tetrad $\tilde{e}_{a}^{\mu}$, so we use Eq.~\eqref{eq:normalized_tetrad} to convert $h_{\tilde{a}\tilde{b}}^{(0,1)}$ to the basis of $e_{a}^{\mu}$.

We compute $h_{\tilde{a}\tilde{b}}^{(0,1)}$ using the {\texttt{Mathematica}} notebooks developed by \cite{Dolan:2023enf}. In \cite{Dolan:2023enf}, the final result of the reconstructed metric $h_{\tilde{a}\tilde{b}}^{(0,1)}$ is decomposed into spin-weighted spherical harmonics ${}_{s}Y_{\ell m}(\theta)$ [instead of spin-weighted spheroidal harmonics ${}_{s}S_{\ell m}(\theta)$], i.e.,
\begin{equation} \label{eq:h_decomp}
    \mathcal{N}^{\tilde{a}\tilde{b}}h_{\tilde{a}\tilde{b}}^{(0,1)}
    =\sum_{\ell_g,m_g}{}_{s}R^{\tilde{a}\tilde{b}}_{\ell_g m_g\omega_g}(r)\,
    {}_{s}Y_{\ell_g m_g}(\theta)e^{-im_g\Omega_gt+im_g\phi}\,,
\end{equation}
where summation over $\tilde{a}$ and $\tilde{b}$ is not implied on the left-hand side, while $\mathcal{N}^{\tilde{a}\tilde{b}}=\mathcal{N}^{\tilde{a}\tilde{b}}(r,\theta)$ is the normalization factor of each component $h_{\tilde{a}\tilde{b}}^{(0,1)}$ chosen in \cite{Dolan:2023enf},
\begin{align}
    & \mathcal{N}^{l_{+}l_{+}}=\mathcal{N}^{l_{-}l_{-}}
    =\mathcal{N}^{m_{+}m_{+}}=\mathcal{N}^{m_{-}m_{-}}=1\,,\nonumber\\
    & \mathcal{N}^{l_{+}m_{+}}
    =\mathcal{N}^{l_{-}m_{-}}=\Gamma(r,\theta)\,,\nonumber\\
    & \mathcal{N}^{l_{+}m_{-}}
    =\mathcal{N}^{l_{-}m_{+}}=\bar{\Gamma}(r,\theta)\,,\nonumber\\
    & \mathcal{N}^{l_{+}l_{-}}
    =\Gamma(r,\theta)\bar{\Gamma}(r,\theta)\Delta(r)\,.
\end{align}
The component $h_{m_{+}m_{-}}^{(0,1)}$ can be obtained from $h_{l_{+}l_{-}}^{(0,1)}$ and the trace $h^{(0,1)}$ of the reconstructed metric $h_{\tilde{a}\tilde{b}}^{(0,1)}$, where
\begin{equation}
    h^{(0,1)}=\frac{\Delta(r)h_{l_{+}l_{-}}^{(0,1)}+h_{m_{+}m_{-}}^{(0,1)}}{\Gamma(r,\theta)\bar{\Gamma}(r,\theta)}\,.
\end{equation}
The constant $\Omega_g$ is the orbital frequency of the secondary and related to the orbital radius $r_0$ of the secondary by
\begin{equation} \label{eq:Omega_g}
    \Omega_{g}=\frac{1}{\sqrt{r_0^{3}/M}+a}\,.
\end{equation}
The spin weight $s$ within the decomposition in Eq.~\eqref{eq:h_decomp} is determined by the number of $m_{+}^{\mu}$ contracted onto $h_{\mu\nu}^{(0,1)}$ minus the number of $m_{-}^{\mu}$ contracted onto the same tensor. For example, the spin weight is $s=\pm2$ for $h_{m_{\pm}m_{\pm}}^{(0,1)}$, respectively. In this work, we include terms up to $\ell_g=18$ in Eq.~\eqref{eq:h_decomp} (to be consistent with \cite{Dyson:2025dlj}) for all the components of $h_{\tilde{a}\tilde{b}}^{(0,1)}$, with ${}_{s}R^{\tilde{a}\tilde{b}}_{\ell_g m_g\omega_g}(r)$ directly obtained from the {\texttt{Mathematica}} notebooks in \cite{Dolan:2023enf}. This truncation in $\ell_g$ results in relative fractional errors $\lesssim 10^{-7}$ when calculating energy fluxes for the values $r_0$ considered here, where we define the relative fractional error/difference of $A$ from $B$ as $\left|1-A/B\right|$.

\begin{figure*}[t]
    \centering
    \includegraphics[width=\linewidth]{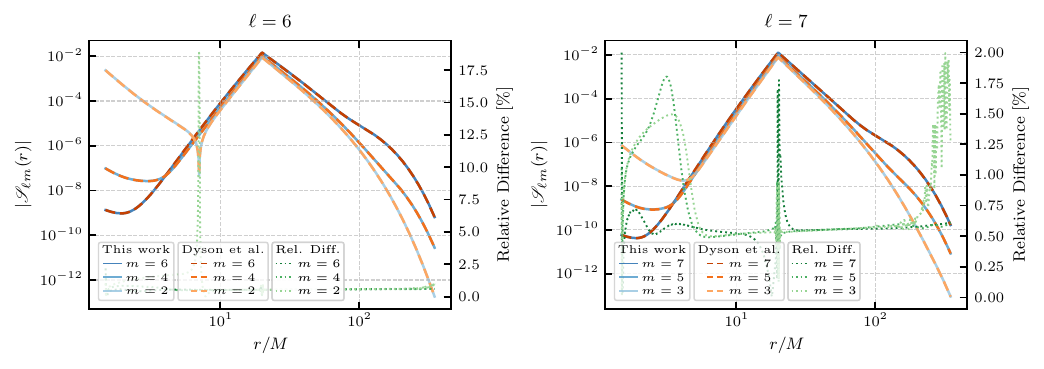}
    \caption{The radial source $\mathscr{S}_{\ell m}(r)$ (left axis) for the same cloud configuration in Fig.~\ref{fig:source_l2l3} at $\ell=6,7$ and $r_0=20M$ for positive $m$. The solid and dashed lines are the sources obtained via the approach in this work and the one in \cite{Dyson:2025dlj}, respectively. The dotted lines show the relative fractional differences in percentages (right axis) between these two approaches in the values of $\left|\mathscr{S}_{\ell m}(r)\right|$.}
    \label{fig:source_l6l7}
\end{figure*}

The derivatives of the quasi-bound state of the scalar cloud $\Phi_{;ab}^{(1,0)}$ can be further expressed in terms of the NP directional derivatives $\left\{D,\mathbf{\Delta},\delta,\bar{\delta}\right\}$ along the tetrad in Eq.~\eqref{eq:normalized_tetrad} and the corresponding spin coefficients. To extract the radial part of Eq.~\eqref{eq:EOM_scalar_11_source}, we follow \cite{Chandrasekhar_1983, Dolan:2023enf, Ma:2024qcv} to replace all NP directional derivatives and spin coefficients with the Chandrasekhar operators $\left\{\mathcal{D}_{n},\mathcal{D}^{\dagger}_{n},
\mathcal{L}_{n},\mathcal{L}^{\dagger}_{n}\right\}$ and the functions $\left\{\Delta(r),\sin{\theta},\Gamma(r,\theta),\bar{\Gamma}(r,\theta)\right\}$, where
\begin{align} \label{eq:chandra_ops}
    & \mathcal{D}_n
    =\mathcal{D}+2n(r-M)/\Delta(r)\,,\nonumber\\ 
    & \mathcal{D}_n^{\dagger}
    =\mathcal{D}^{\dagger}+2n(r-M)/\Delta(r)\,,\nonumber\\
    & \mathcal{L}_n
    =\mathcal{L}+n\cot\theta\,,\;
    \mathcal{L}_n^{\dagger}
    =\mathcal{L}^{\dagger}+n\cot\theta\,.
\end{align}
The radial operators $(\mathcal{D},\mathcal{D}^{\dagger})$ are directional derivatives along $l_{+}^{\mu}$ and $l_{-}^{\mu}$ in Eq.~\eqref{eq:tetrad}, while the angular operators $(\mathcal{L},\mathcal{L}^{\dagger})$ are directional derivatives along $m_{+}^{\mu}$ and $m_{-}^{\mu}$ in Eq.~\eqref{eq:tetrad}, respectively. In this case, one can apply the commutation relations between these operators and functions in \cite{Chandrasekhar_1983, Dolan:2023enf, Ma:2024qcv} and Appendix~\ref{appendix:chandrasekhar_operators} to reduce $\mathcal{S}_{\Phi}^{(1,1)}$ into a sum of terms of the following form:
\begin{align} \label{eq:S_structure}
    & e^{-i(\omega_c+m_g\Omega_g)t+i(m_c+m_g)\phi} 
    \Gamma^{-\beta}(r,\theta)\bar{\Gamma}^{-\sigma}(r,\theta) \nonumber\\
    & \times\left\{\Delta^{\alpha}(r)\hat{\mathcal{O}}_{r}
    \left[{}_{0}R_{\ell_c m_c\omega_c}^{\Phi}(r)\right]
    {}_{s}R^{\tilde{a}\tilde{b}}_{\ell_g m_g\omega_g}(r)\right\} \nonumber\\
    & \times\left\{\sin^{\kappa}{\theta}\hat{\mathcal{O}}_{\theta}
    \left[{}_{0}S^{\Phi}_{\ell_c m_c\omega_c}(\theta)\right]
    {}_{s}Y_{\ell_g m_g}(\theta)\right\}\,,
\end{align}
where $\alpha,\beta,\kappa,\sigma$ are non-negative integers, and the complete expression of $\mathcal{S}_{\Phi}^{(1,1)}$ is provided in Appendix~\ref{appendix:source_terms}. The operator $\hat{\mathcal{O}}_{r}$ is a function of the radial operators $\mathcal{D}_{n}$ and $\mathcal{D}^{\dagger}_{n}$, while the operator $\hat{\mathcal{O}}_{\theta}$ is a function of the angular operators $\mathcal{L}_{n}$ and $\mathcal{L}^{\dagger}_{n}$. 

Due to the structure of the source term in Eq.~\eqref{eq:S_structure}, the scalar radiation field $\Phi^{(1,1)}$ is decomposed as
\begin{align} \label{eq:phi_11_decomp}
    & \Phi^{(1,1)}=\sum_{\ell, m}{}_{0}\mathscr{R}^{\Phi}_{\ell m\omega}(r)\,
    {}_{0}S_{\ell m\omega}(\theta)e^{-i\omega t+im\phi}\,, \nonumber\\
    & m=m_c+m_g\,,\quad\omega=\omega_c+m_g\Omega_g\,.
\end{align}
To extract the radial part of Eq.~\eqref{eq:EOM_scalar_11}, one can then integrate it against ${}_{0}\bar{S}_{\ell m\omega}(\theta)e^{-im\phi}$ over the 2-sphere and use the orthogonality relation of ${}_{s}S_{\ell m\omega}(\theta)$, i.e.,
\begin{equation} \label{eq:SWSH_orth}
    \int_{0}^{\pi} {}_{s}S_{\ell m\omega}(\theta)
    {}_{s}\bar{S}_{\ell' m\omega}(\theta)
    \sin{\theta}\,d{\theta}
    =\frac{1}{2\pi}\delta_{\ell\ell'}\,.
\end{equation}

One challenge when extracting the radial part of terms in the form of Eq.~\eqref{eq:S_structure} is that $\beta\geq0$ and $\sigma\geq0$ for all terms in this work, so one has to evaluate the angular projection numerically at each radial point on a discretized radial grid. This process can be computationally expensive, so we instead choose to apply the method in \cite{Spiers:2024src} by decomposing factors of $\Gamma(r,\theta)$ and $\bar{\Gamma}(r,\theta)$ into Fourier series of $\theta$. More specifically, we first express the factor $\Gamma^{-\beta}(r,\theta)\bar{\Gamma}^{-\sigma}(r,\theta)$ as
\begin{equation} \label{eq:gamma_regroup}
     \Gamma^{-\beta}(r,\theta)\bar{\Gamma}^{-\sigma}(r,\theta)
     =\begin{cases}
     \Sigma^{-\beta}(r,\theta)\bar{\Gamma}^{\beta-\sigma}(r,\theta) 
     & \text{if } \beta \geq \sigma\\
     \Sigma^{-\sigma}(r,\theta)\Gamma^{\sigma-\beta}(r,\theta) 
     & \text{if } \beta < \sigma\,,
     \end{cases}
\end{equation}
where $\Sigma(r,\theta)=\Gamma(r,\theta)\bar{\Gamma}(r,\theta)=r^2+a^2\cos^2{\theta}$. Since now the exponent of $\Gamma(r,\theta)$ or $\bar{\Gamma}(r,\theta)$ is positive, we can expand them via the binomial theorem, i.e.,
\begin{align}
    (r\pm ia\cos{\theta})^n=
    \sum_{k=0}^{n}r^{n-k}(\pm ia\cos{\theta})^k\,,
\end{align}
where $n=|\beta-\sigma|$. For the factor of $\Sigma(r,\theta)$, we then perform a Fourier series decomposition
\begin{equation} \label{eq:sigma_expansion}
    \Sigma^{-\beta}(r,\theta)=\sum_{p=0}^{\infty}f_p(r)\cos (p\theta) \,,
\end{equation}
where we have used the fact that $\Sigma(r,\theta)$ is a periodic and even function of $\theta$. Since $\Sigma(r,\theta)$ only contains even powers of $\cos{\theta}$, $f_p(r)=0$ when $p$ is odd. The radial function $f_p(r)$ can be determined from
\begin{align}
    & f_0(r)=\frac{1}{2\pi}\int_{-\pi}^{\pi}
    \Sigma^{-\beta}(r,\theta)\,d\theta\,,\nonumber\\
    & f_p(r)=\frac{1}{\pi}\int_{-\pi}^{\pi}
    \Sigma^{-\beta}(r,\theta)\cos (p\theta )\,d\theta\,.
\end{align}
As demonstrated in \cite{Spiers:2024src}, this Fourier series representation of $\Gamma^{-\beta}(r,\theta)\bar{\Gamma}^{-\sigma}(r,\theta)$ can be very accurate if enough terms are included in Eq.~\eqref{eq:sigma_expansion} and enough precision is prescribed for $r$ when evaluating $f_p(r)$. In practice, the sum in Eq.~\eqref{eq:sigma_expansion} is truncated at some value $p=p_{\max}$. For the purpose of LISA data analysis, sufficient accuracy is achieved for $p_{\max}\gtrsim 2\beta$ \cite{Spiers:2024src}. In our case, since $0\leq\max(\beta,\sigma)\leq3$, we set $p_{\max}=12$ and use $64$ digits of precision for $r$ when evaluating $f_p(r)$. All radial coefficients $f_p(r)$ are pre-computed before projecting Eq.~\eqref{eq:S_structure} onto the 2-sphere. Doing so, one can reduce Eq.~\eqref{eq:S_structure} into a sum of terms of the form
\begin{align} \label{eq:S_structure_2}
    & e^{-i(\omega_c+m_g\Omega_g)t+i(m_c+m_g)\phi} \nonumber\\
    & \times\left\{r^{|\beta-\sigma|-k}f_p(r)\Delta^{\alpha}(r)\hat{\mathcal{O}}_{r}
    \left[{}_{0}R_{\ell_c m_c\omega_c}^{\Phi}(r)\right]
    {}_{s}R^{\tilde{a}\tilde{b}}_{\ell_g m_g\omega_g}(r)\right\} \nonumber\\
    & \times\left\{\cos^k{\theta}\cos{p\theta}
    \sin^{\kappa}{\theta} \; \hat{\mathcal{O}}_{\theta}
    \left[{}_{0}S^{\Phi}_{\ell_c m_c\omega_c}(\theta)\right]
    {}_{s}Y_{\ell_g m_g}(\theta)\right\}\,,
\end{align}
up to some constants. Notice that the second and third lines of Eq.~\eqref{eq:S_structure_2} are now purely radial and purely angular, respectively, so we can easily extract the radial part.

Using the orthogonality relation in Eq.~\eqref{eq:SWSH_orth}, one can then integrate Eq.~\eqref{eq:EOM_scalar_11}, with $\mathcal{S}_{\Phi}^{(1,1)}$ being a sum of terms in the form of Eq.~\eqref{eq:S_structure_2}, against the spin-$0$ spheroidal harmonics over the 2-sphere to extract its radial part. In the end, we find
\begin{align} \label{eq:EOM_scalar_11_radial}
    & \bigg[\frac{d}{d r}\left(\Delta(r)\frac{d}{dr}\right) 
     +\frac{\omega^2\left(r^2+a^2\right)^2-4Mam\omega r+a^2m^2}{\Delta(r)} \nonumber\\
    & -\left(a^2\omega^2+\mu^2r^2+\Lambda_{\ell m}\right)\bigg]
    {}_{0}\mathscr{R}^{\Phi}_{\ell m\omega}(r)
    =\mathscr{S}_{\ell m}(r)\,,
\end{align}
where $m$ and $\omega$ are given by the selection rule in Eq.~\eqref{eq:phi_11_decomp}. The radial source term $\mathscr{S}_{\ell m}(r)$ is a sum of terms in the form of
\begin{equation} \label{eq:radial_source}
    \mathscr{C}_{\ell m}
    r^{|\beta-\sigma|-k}f_p(r)\Delta^{\alpha}(r)\hat{\mathcal{O}}_{r}
    \left[{}_{0}R_{\ell_c m_c\omega_c}^{\Phi}(r)\right]
    {}_{s}R^{\tilde{a}\tilde{b}}_{\ell_g m_g\omega_g}(r)\,,
\end{equation}
up to some constants in $M$ and $a$, where the angular projection coefficient $\mathscr{C}_{\ell m}$ comes from
\begin{align} \label{eq:angular_projection}
    \mathscr{C}_{\ell m}
    =& \;2\pi\int_{0}^{\pi}\Big\{\cos^k{\theta}\cos{p\theta}\sin^{\kappa}{\theta}
    \hat{\mathcal{O}}_{\theta}\left[{}_{0}S^{\Phi}_{\ell_c m_c\omega_c}(\theta)\right]\nonumber\\
    & \;{}_{s}Y_{\ell_g m_g}(\theta){}_{0}\bar{S}_{\ell m\omega}(\theta)
    \Big\}\sin{\theta}d\theta\,.
\end{align}
Since one needs to re-evaluate $\mathscr{S}_{\ell m}(r)$ at different orbital radius $r_0$, we choose not to provide the complete expression of $\mathscr{S}_{\ell m}(r)$ here. Nonetheless, one can easily obtain $\mathscr{S}_{\ell m}(r)$ by applying the procedures above to the source $S^{\ell m(1,1)}_{\Phi}$ provided in Appendix~\ref{appendix:source_terms}. Specifically, one can replace the factors of $\Gamma(r,\theta)$ and $\bar{\Gamma}(r,\theta)$ in Eq.~\eqref{eq:source_components} with Eq.~\eqref{eq:gamma_regroup} and decompose them using the expansion in Eq.~\eqref{eq:sigma_expansion}. Then, one can extract the radial part $\mathscr{S}_{\ell m}(r)$ of the source term by integrating Eq.~\eqref{eq:source_complete} against ${}_{0}\bar{S}_{\ell m\omega}(\theta)e^{-im\phi}$ over the 2-sphere [i.e., Eq.~\eqref{eq:angular_projection}]. The \texttt{Mathematica} notebook implementing these procedures will be provided upon request.

The radial source $\mathscr{S}_{\ell m}(r)$ obtained via this procedure has two main sources of errors: the truncation of the $(\ell_g,m_g)$ modes when reconstructing the metric perturbation $h_{\mu\nu}^{(0,1)}$ [i.e., Eq.~\eqref{eq:h_decomp}] and the truncation of the Fourier series at order $p_{\max}$ in Eq.~\eqref{eq:sigma_expansion} when evaluating the factors of $\Gamma(r,\theta)$ and $\bar{\Gamma}(r,\theta)$. For the former, we have verified that a truncation at $\ell_g=18$ induces relative fractional errors $\lesssim10^{-7}$ in both the horizon and infinity fluxes at the chosen $r_0$ we tested. For the latter, we found that if we choose $p_{\max}=12$ and use $64$ digits of precision for $r$, the relative fractional errors in $\Sigma(r,\theta)^{-\beta}$ with $0\leq\beta\leq3$ are $\lesssim 10^{-6}$ near the horizon and decrease exponentially at larger $r$. Other minor errors include the truncation of the continuous fraction when using Leaver's method to solve for ${}_{0}R_{\ell_c m_c}^{\Phi}(r)$ and other intrinsic errors when calculating the radial part ${}_{s}R^{\tilde{a}\tilde{b}}_{\ell_g m_g}(r)$ of the reconstructed metric $h_{\mu\nu}^{(0,1)}$ using the \texttt{Mathematica} notebooks in \cite{Dolan:2023enf}.

\begin{figure*}[t]
    \centering
    \includegraphics[width=\linewidth]{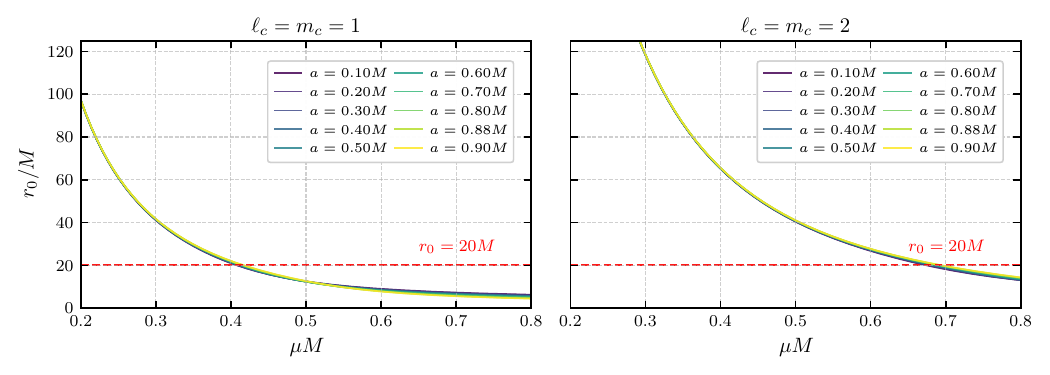}
    \caption{The threshold $r_0$ in Eq.~\eqref{eq:threshold_r0} versus the scalar mass $\mu$, across which the radial part ${}_{0}\mathscr{R}_{\ell m\omega}^{\Phi}(r)$ of the scalar radiation $\Phi^{(1,1)}$ transitions from an $1/r$ decay to an exponential decay at infinity. The left and right subplots are for the $(\ell_c=m_c=1,n_c=0)$ and $(\ell_c=m_c=2,n_c=0)$ clouds, respectively. Here, we have set $m_g=1$, and different curves are for BHs of different spins $a$. The red horizontal line marks the threshold $r_0$ at $20M$.} 
    \label{fig:r_threshold}
\end{figure*}

In order to study the accuracy of the radial source $\mathscr{S}_{\ell m}(r)$, we have compared it to the results in \cite{Dyson:2025dlj} at $r_0=20M$ for the $\ell_c=m_c=1$ cloud. Similar to this work, Ref.~\cite{Dyson:2025dlj} first calculates the background scalar field $\Phi^{(1,0)}$ using Leaver's method in \cite{Leaver:1985ax, Dolan:2007mj} and then contracts the derivatives of $\Phi^{(1,0)}$ with the reconstructed metric $h_{\mu\nu}^{(0,1)}$ from \cite{Dolan:2023enf} to construct the source in Eq.~\eqref{eq:EOM_scalar_11_source_Lorentz}. However, instead of projecting out the radial part of the source semi-analytically as discussed in this section, Ref.~\cite{Dyson:2025dlj} evaluates the source on a 2-$d$ numerical grid and then integrates this purely numerical 2-$d$ source against spin-weighted spheroidal harmonics to extract $\mathscr{S}_{\ell m}(r)$. Thus, these two different approaches serve as good independent checks of each other, although the semi-analytical simplifications in this work might prove helpful when dealing with the much more complicated source and angular projection when calculating gravitational fluxes. In Figs.~\ref{fig:source_l2l3} and \ref{fig:source_l6l7}, we present $\mathscr{S}_{\ell m}(r)$ at $r_0=20M$ computed via these two approaches for the $\ell=2,3$ modes and the $\ell=6,7$ modes, respectively. As discussed in \cite{Dyson:2025dlj}, due to selection rules, the source is zero when $\ell$ and $m$ are of opposite parity [i.e., $\ell$ and $m$ are even (odd) and odd (even), respectively], so we do not present these modes in Figs.~\ref{fig:source_l2l3} and \ref{fig:source_l6l7}. Both figures show good agreement between sources obtained using these two independent methods, with relative fractional differences $\lesssim 3\%$ across most values of $r$, even for the subdominant modes. The only exception is when $\ell=6,m=2$ at $r\sim7M$, where the source is more singular due to the reconstructed metric. We have also found agreement for other modes and values of $r_0$. In the next section, we will solve Eq.~\eqref{eq:EOM_scalar_11_radial} via a Green's function method to obtain the scalar radiation $\Phi^{(1,1)}$.

\section{The scalar radiation}
\label{sec:scalar_radiation}

\begin{figure*}[t]
    \centering
    \includegraphics[width=0.8\linewidth]{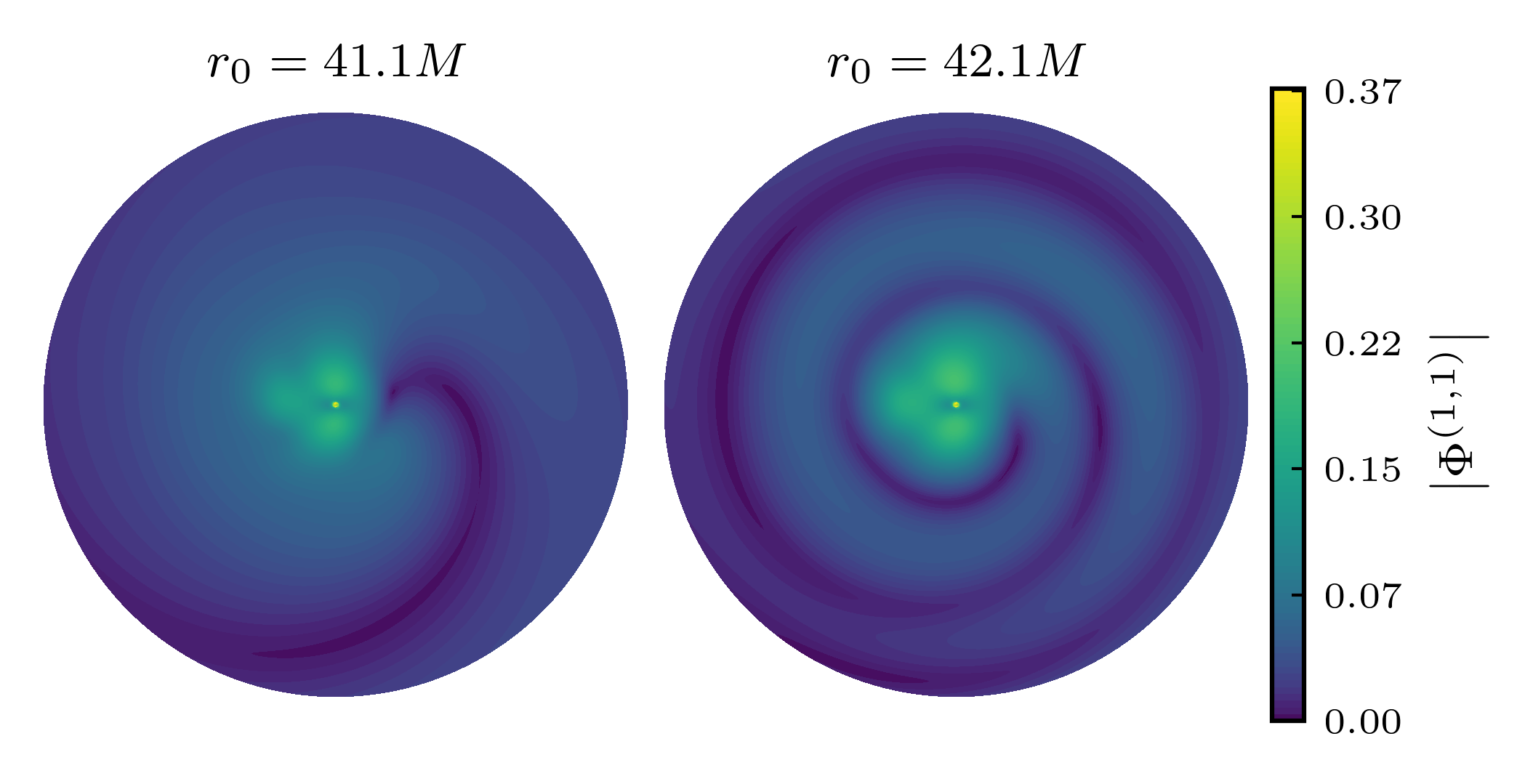}
    \caption{The scalar radiation $\Phi^{(1,1)}$ of the $(\ell_c=m_c=1,n_c=0)$ scalar cloud around a Kerr BH of spin $a=0.88M$ with scalar mass $\mu M=0.3$ on the $(r,\phi)$ plane at $t=0$. We have summed up all the $(\ell,m)$ modes with $2\leq\ell\leq5$. The left and right subplots are for the secondary at $r_0=41.1M$ and $r_0=42.1M$, respectively. The scalar radiation shown above is at the equilateral plane (i.e., $\theta=0$) up to the radius $r=200M$, where the $z$-axis of the coordinate aligns with the BH spin.} 
    \label{fig:scalar_radiation_11}
\end{figure*}

In this section, we solve Eq.~\eqref{eq:EOM_scalar_11_radial} with the source term $\mathscr{S}_{\ell m}(r)$ given by the sum of terms in the form of Eq.~\eqref{eq:radial_source}. Following \cite{Teukolsky:1973ha}, we first rewrite Eq.~\eqref{eq:EOM_scalar_11_radial} in the tortoise coordinate $r_*$, defined via
\begin{equation}
    dr_{*}/dr=(r^2+a^2)/\Delta(r)\,,
\end{equation}
and in terms of the radial function ${}_{0}\hat{\mathscr{R}}^{\Phi}_{\ell m\omega}(r)$, where
\begin{equation}
    {}_{0}\hat{\mathscr{R}}^{\Phi}_{\ell m\omega}(r)
    =\sqrt{r^2+a^2}\,{}_{0}\mathscr{R}^{\Phi}_{\ell m\omega}(r)\,,
\end{equation}
such that Eq.~\eqref{eq:EOM_scalar_11_radial} becomes
\begin{align} \label{eq:EOM_scalar_11_radial_r*}
    & \Big\{\partial_{r_{*}}^2+\left[K^2(r)-\left(\lambda_{\ell m}+\mu^2r^2\right)\Delta(r)\right]/\left(r^2+a^2\right)^2-G^2(r) \nonumber\\
    & -\partial_{r_{*}}G(r)\Big\} 
    {}_{0}\hat{\mathscr{R}}^{\Phi}_{\ell m\omega}(r)
    =\mathscr{S}_{\ell m}(r)\Delta(r)/(r^2+a^2)^{3/2}\,,
\end{align}
where $K(r)=(r^2+a^2)\omega-am$, $\lambda_{\ell m}=\Lambda_{\ell m}+a^2\omega^2-2am\omega$, and $G(r)=r\Delta(r)/(r^2+a^2)^2$. In the limit $r\rightarrow\infty$ or $r_{*}\rightarrow\infty$, the homogeneous part of Eq.~\eqref{eq:EOM_scalar_11_radial_r*} becomes
\begin{equation}
    \left[\partial_{r_{*}}^2+\left(\omega^2-\mu^2+2M\mu^2/r\right)
    +\mathcal{O}(r^{-2})\right]
    {}_{0}\hat{\mathscr{R}}^{\Phi}_{\ell m\omega}(r)=0\,,
\end{equation}
while in the limit $r\rightarrow r_{+}$ or $r_{*}\rightarrow-\infty$, the homogeneous part of Eq.~\eqref{eq:EOM_scalar_11_radial_r*} becomes
\begin{equation}
    \left[\partial_{r_{*}}^2+\left(\omega-m\omega_{+}\right)^2
    +\mathcal{O}(r)\right]
    {}_{0}\hat{\mathscr{R}}^{\Phi}_{\ell m\omega}(r)=0\,,
\end{equation}
where $\omega_{+}=a/(2Mr_{+})$ is the horizon frequency. Thus, we can construct two independent homogeneous solutions to Eq.~\eqref{eq:EOM_scalar_11_radial}, ${}_{0}\mathscr{R}_{\ell m\omega}^{\textrm{in}}(r)$ and ${}_{0}\mathscr{R}_{\ell m\omega}^{\textrm{up}}(r)$, that satisfy the boundary conditions \cite{Teukolsky:1973ha, Dolan:2007mj, Dyson:2025dlj}
\begin{subequations} \label{eq:scalar_boundary}
\begin{align}
    & {}_{0}\mathscr{R}_{\ell m\omega}^{\textrm{in}}(r\rightarrow r_+)
    \sim (r-r_{+})^{\frac{-2iMr_{+}(\omega-\omega_{+})}{r_{+} - r_{-}}}\,, \\
    & {}_{0}\mathscr{R}_{\ell m\omega}^{\textrm{up}}(r\rightarrow\infty)
    \sim \frac{e^{i\sqrt{\omega^2-\mu^2}r_{*}}r^{iM\mu^2/\sqrt{\omega^2-\mu^2}}}{\sqrt{r^2+a^2}}\,,
\end{align}  
\end{subequations}
respectively. Following \cite{Pound:2021qin}, we can then construct the solution to Eq.~\eqref{eq:EOM_scalar_11_radial} via Green's function as 
\begin{widetext}
\begin{equation} \label{eq:scalar_sol_11}
    {}_{0}\mathscr{R}_{\ell m\omega}^{\Phi}(r)
    =\frac{{}_{0}\mathscr{R}_{\ell m\omega}^{\textrm{up}}(r)
    \int_{r_{+}}^{r}{}_{0}\mathscr{R}_{\ell m\omega}^{\textrm{in}}(r')
    \mathscr{S}_{\ell m}(r')dr'
    +{}_{0}\mathscr{R}_{\ell m\omega}^{\textrm{in}}(r)\int_{r}^{\infty}
    {}_{0}\mathscr{R}_{\ell m\omega}^{\textrm{up}}(r')
    \mathscr{S}_{\ell m}(r')dr'}
    {\Delta(r)\left({}_{0}\mathscr{R}_{\ell m\omega}^{\textrm{in}}(r)
    \partial_{r}{}_{0}\mathscr{R}_{\ell m\omega}^{\textrm{up}}(r)
    -{}_{0}\mathscr{R}_{\ell m\omega}^{\textrm{up}}(r)
    \partial_{r}{}_{0}\mathscr{R}_{\ell m\omega}^{\textrm{in}}(r)\right)}\,,
\end{equation}    
\end{widetext}
where we solve for ${}_{0}\mathscr{R}_{\ell m\omega}^{\textrm{in}}(r)$ and ${}_{0}\mathscr{R}_{\ell m\omega}^{\textrm{up}}(r)$ by numerically integrating their respective asymptotic expansions from the horizon and infinity toward infinity and the horizon using Eq.~\eqref{eq:EOM_scalar_11_radial}, respectively. To more accurately capture the boundary conditions near the horizon and infinity, especially when the scalar radiation's wavelength becomes much longer as $\omega^2-\mu^2\rightarrow0$, we asymptotically expand the homogeneous solutions ${}_{0}\mathscr{R}_{\ell m\omega}^{\textrm{in}}(r)$ about the horizon and $ {}_{0}\mathscr{R}_{\ell m\omega}^{\textrm{up}}$ about spatial infinity using Eq.~\eqref{eq:scalar_boundary} as
\begin{subequations} \label{eq:scalar_boundary_expand}
\begin{align}
    & {}_{0}\mathscr{R}_{\ell m\omega}^{\textrm{in}}(r)
    =(r-r_{+})^{\frac{-2iMr_{+}(\omega-\omega_{+})}{r_{+} - r_{-}}}
    \sum_{j=0}^{\infty}A_j(r-r_{+})^j\,, \\
    & {}_{0}\mathscr{R}_{\ell m\omega}^{\textrm{up}}(r)
    =\frac{e^{i\sqrt{\omega^2-\mu^2}r_{*}}
    r^{iM\mu^2/\sqrt{\omega^2-\mu^2}}}{\sqrt{r^2+a^2}}
    \sum_{j=0}^{\infty}B_jr^{-j}\,,
\end{align}     
\end{subequations}
where $A_0=B_0=1$, and we include terms up to $j=4$. The coefficients $A_j$ and $B_j$ can be found by inserting Eq.~\eqref{eq:scalar_boundary_expand} into the left-hand side of Eq.~\eqref{eq:EOM_scalar_11_radial} and solving the equation order-by-order in $(r-r_{+})$ and $1/r$, respectively. Their complete expressions for $1\leq j\leq4$ are provided in the supplementary notebook \cite{boundary_coeffs}.

\begin{figure*}[t]
    \centering
    \includegraphics[width=0.8\linewidth]{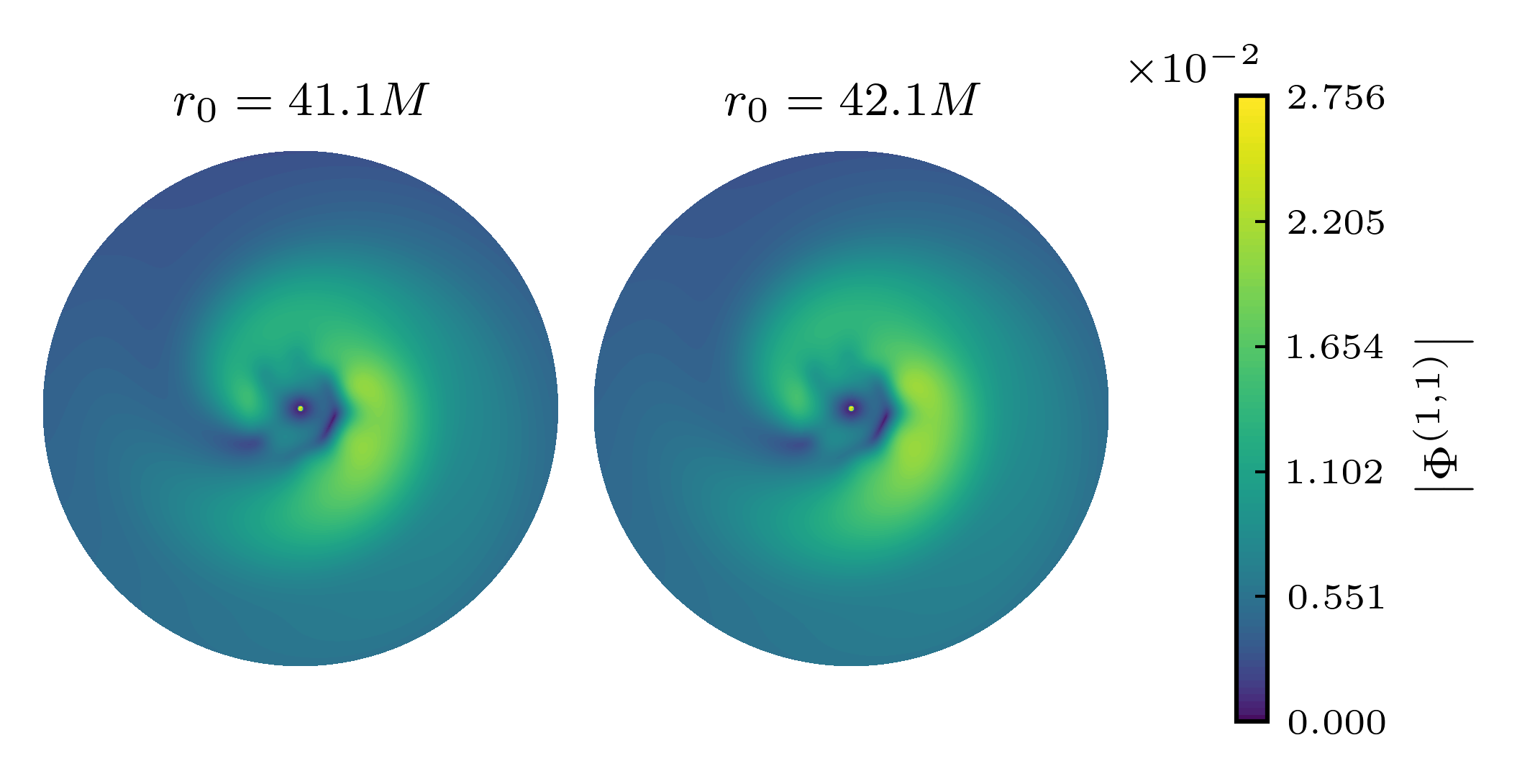}
    \caption{The scalar radiation $\Phi^{(1,1)}$ of the $(\ell_c=m_c=2,n_c=0)$ scalar cloud around a Kerr BH of spin $a=0.88M$ with scalar mass $\mu M=0.3$ on the $(r,\phi)$ plane at $t=0$. We have summed up all the $(\ell,m)$ modes with $3\leq\ell\leq5$ and $-3\leq m\leq5$. The left and right subplots are for the secondary at $r_0=41.1M$ and $r_0=42.1M$, respectively. The scalar radiation shown above is at the equilateral plane (i.e., $\theta=0$) up to the radius $r=200M$, where the $z$-axis of the coordinate aligns with the BH spin.} 
    \label{fig:scalar_radiation_22}
\end{figure*}

One notable feature of $\left|{}_{0}\mathscr{R}_{\ell m\omega}^{\textrm{up}}(r)\right|$ near infinity [i.e., Eq.~\eqref{eq:scalar_boundary}] is that this function transitions from a $1/r$ decay, when $|\omega|>\mu$, to an exponential decay at large $r$, when $|\omega|<\mu$, or more simply when
\begin{equation} 
    \Omega_g<\frac{\mu-\textrm{sgn}(m_g)\omega_c}{|m_g|}\,.
\end{equation}
Recall that $\omega=\omega_c+m_g\Omega_g$ in Eq.~\eqref{eq:phi_11_decomp}, so $|\omega|<\mu$ at $m_g=0$ always. The above arguments used that $0\leq(\mu-\omega_c)M<1$ for a scalar cloud in the quasibound state with $\mu M<1$. Given Eq.~\eqref{eq:scalar_sol_11}, the radial part ${}_{0}\mathscr{R}_{\ell m\omega}^{\Phi}(r)$ of the scalar radiation $\Phi^{(1,1)}$ will experience the same transition. When a certain mode of $\Phi^{(1,1)}$ decays exponentially at infinity, this mode cannot contribute to the infinity flux. Using the relation between $\Omega_g$ and $r_0$ in Eq.~\eqref{eq:Omega_g}, we then find that this transitional behavior occurs when
\begin{equation} \label{eq:threshold_r0}
    r_0>M^{1/3}\left(\frac{|m_g|}{\mu-\textrm{sgn}(m_g)\omega_c}
    -a\right)^{2/3}\,.
\end{equation}
When $m_g>0$, the smallest orbital radius $r_0$ where such a transition occurs is when $m_g=1$. This threshold $r_0$ for a $(\ell_c=m_c=1,n_c=0,\mu M=0.3)$ cloud around a Kerr BH of spin $a=0.88M$, with $\omega_c$ given in Eq.~\eqref{eq:omega_c_11}, is $r_0\approx41.66M$, as also found in \cite{Dyson:2025dlj}. As we increase $r_0$ beyond this threshold, higher $m_g$ modes start to decay exponentially at large $r$ in sequence (e.g., the threshold $r_0$ for $m_g=2$ is $r_0\approx66.20M$) and do not contribute to the energy flux at spatial infinity. In contrast, when $m_g<0$, the smallest transition radius $r_0$ is much lower, with $r_0\approx0.86M$ at $m_g=-1$, given the BH spin $a=0.88M$, the scalar mass $\mu M=0.3$, and the scalar cloud's frequency $\omega_c$ in Eq.~\eqref{eq:omega_c_11}. Since $r_0=0.86M$ lies inside the horizon, the scalar radiation driven by the $m_g=-1$ mode does not contribute to the energy flux at infinity. This threshold radius also increases when $m_g$ becomes more negative, reaching $r_0\approx 4.38M$ at $m_g=-6$, which is the most negative mode considered in this work. Thus, we can ignore negative $m_g$ modes at most $r_0$ when calculating the infinity flux. 

In Fig.~\ref{fig:r_threshold}, we plot the threshold $r_0$ of Eq.~\eqref{eq:threshold_r0} at $m_g=1$ against the scalar mass $\mu$ for several BH spins $a\in[0.2,0.9]M$. LISA is sensitive to EMRIs emitting GWs with frequencies as low as $\sim 1\,\mathrm{mHz}$ \cite{LISA:2024hlh}, which corresponds to an orbital radius $r_0\approx20M$ when the central supermassive BH is of mass $M=10^6M_{\odot}$ and spin $a=0.88M$. Thus, for the $\ell_c=m_c=1$ cloud with scalar mass $\mu M=0.3$ and frequency $\omega_c$ in Eq.~\eqref{eq:omega_c_11} around such a BH, the transition at $r_0\approx41.66M$ is not observable. Nonetheless, in Fig.~\ref{fig:r_threshold}, we notice that when the scalar mass $\mu M\gtrsim 0.4$, this threshold $r_0$ falls below $20M$, possibly leading to an observable signature. As $\mu M$ approaches $1$, the treatment in this work starts to break down since the expansion parameter $\zeta\propto(\mu M)^3$ becomes large and the cloud may no longer be a quasi-bound state. For clouds of higher $\ell_c$, such as the $\ell_c=m_c=2$ cloud shown in Fig.~\ref{fig:r_threshold}, coupled to GWs of higher $m_g$, this threshold $r_0$ only drops below $20M$ for very large $\mu M$, so this transition may not be observable. Furthermore, we also observe that this threshold $r_0$ depends weakly on the BH spin. This is because the first term inside the parenthesis of Eq.~\eqref{eq:threshold_r0} dominates over $a$, and the frequency $\omega_c$ depends weakly on $a$ for small $\mu M$ \cite{Baumann:2018vus}.

The transition discussed above can be directly seen in Fig.~\ref{fig:scalar_radiation_11}, where we plot the scalar radiation $\Phi^{(1,1)}$ driven by the dipolar cloud in the equilateral plane up to $r=200M$ for a secondary at $r_0=41.1M$ and $r_0=42.1M$, respectively. To compare our results with the ones in \cite{Dyson:2025dlj}, we have summed up the $(\ell,m)$ modes that contribute to the infinity flux, with $2\leq\ell\leq5$. As discussed above, the modes with $\ell=0,1$ do not contribute to the infinity flux, so they are not included in Fig.~\ref{fig:scalar_radiation_11}. Similar to \cite{Dyson:2025dlj}, we observe a qualitative change in the scalar radiation at large $r$ when the secondary moves across the threshold orbital radius $r_0\approx41.66M$. This qualitative change is largely due to the transition of the $m_g=1$ mode's behavior at large $r$ discussed above, so if one manually removes the contribution of the $m_g=1$ mode from Fig.~\ref{fig:scalar_radiation_11}, the two subplots will become almost identical.

In Fig.~\ref{fig:scalar_radiation_22}, we also plot the scalar radiation $\Phi^{(1,1)}$ driven by the quadrupolar cloud, including the modes with $3\leq\ell\leq5$ and $-3\leq m\leq5$. For the same reason discussed above, the modes with $0\leq\ell\leq2$ do not contribute to the infinity flux. For this cloud, the scalar radiation driven by the $m_g=1$ mode of GWs decays exponentially at large $r$ when $r_0\gtrsim 74.63M$. Since this radius is outside the frequency band of LISA for most of the EMRIs \cite{LISA:2024hlh}, we choose to plot the scalar radiation at $r_0=41.1M$ and $r_0=42.1M$ as a comparison to Fig.~\ref{fig:scalar_radiation_11}. As expected, the scalar radiation profile changes little across $r_0 = 41.66M$, but differs significantly from that of the dipolar cloud. For equal cloud mass, the quadrupolar cloud produces much weaker and more spatially extended radiation, reflecting its broader profile and greater separation from the horizon compared to the dipolar case, as shown in Fig.~\ref{fig:scalar_profile}.

\section{Energy and angular momentum fluxes}
\label{sec:fluxes}

In this section, we compute the energy and angular momentum fluxes associated with the scalar radiation $\Phi^{(1,1)}$ following the prescription in \cite{Brito:2023pyl, Dyson:2025dlj}. The change of the secondary's orbital energy $\dot{E}_{\mathrm{orb}}$ and angular momentum $\dot{L}_{\mathrm{orb}}$ has contributions from both the scalar and gravitational radiations \cite{Dyson:2025dlj},
\begin{subequations}
\begin{align} \label{eq:energy_balance}
    \dot{E}_{\mathrm{orb}}
    &=-\dot{M}_{c}-\dot{E}^{\Phi,\infty}-\dot{E}^{\Phi,H}
    -\dot{E}^{h,\infty}-\dot{E}^{h,H}\,, \\
    \dot{L}_{\mathrm{orb}}
    &=-\dot{S}_{c}-\dot{L}^{\Phi,\infty}-\dot{L}^{\Phi,H}
    -\dot{L}^{h,\infty}-\dot{L}^{h,H}\,,
\end{align}
\end{subequations}
where $\dot{E}^{\Phi,H/\infty}$ are the horizon and infinity energy fluxes associated with the scalar field's stress-energy tensor $T_{\mu\nu}^{\Phi}$ in Eq.~\eqref{eq:stress_Phi}, and similarly for the angular momentum fluxes $\dot{L}^{\Phi,H/\infty}$. As shown in \cite{Teukolsky:1973ha}, $\dot{E}^{\Phi,H/\infty}$ and $\dot{L}^{\Phi,H/\infty}$ can be computed from
\begin{subequations}
\begin{align}
    & \dot{E}^{\Phi,\infty} 
    =-\lim_{r\rightarrow+\infty}r^2
    \int T_{\mu r}^{\Phi}\xi_{(t)}^\mu\,d\Omega \,, \\
    & \dot{E}^{\Phi,H} 
    =\lim_{r\rightarrow r_{+}}2Mr_{+}
    \int T_{\mu\nu}^{\Phi}\xi_{(t)}^\mu l^\nu\,d\Omega\,, \\
    & \dot{L}^{\Phi,H/\infty}
    =-\dot{E}^{\Phi,H/\infty}
    (\xi_{(t)}^\mu\rightarrow\xi_{(\phi)}^\mu)\,,
\end{align}    
\end{subequations}
where $\xi_{(t)}^\mu=\partial_t$ and $\xi_{(\phi)}^\mu=\partial_{\phi}$ are Killing vectors of the supermassive BH (Kerr background) metric, and so is $l^{\mu}=\xi_{(t)}^\mu+\omega_{+}\xi_{(\phi)}^\mu$ as well. On the other hand, $\dot{E}^{h,H/\infty}$ and $\dot{L}^{h,H/\infty}$ are the energy and angular momentum fluxes associated with gravitational radiation, respectively.

\begin{figure}[t]
    \centering
    \includegraphics[width=\linewidth]{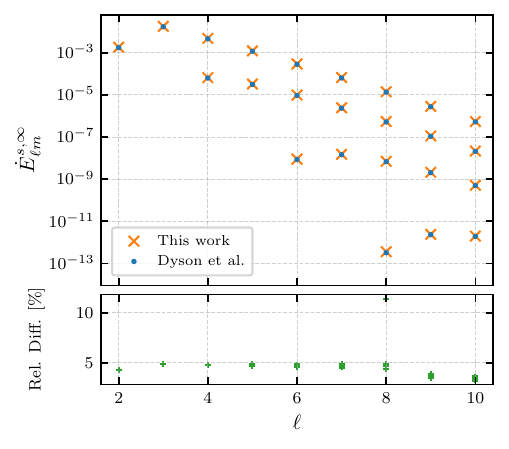}
    \caption{The mode by mode infinity energy flux $\dot{E}_{\ell m}^{s,\infty}$ associated with the scalar radiation $\Phi^{(1,1)}$ of the $(\ell_c=m_c=1,n_c=0)$ scalar cloud around a Kerr BH of spin $a=0.88M$. The scalar field is of mass $\mu M=0.3$, and the secondary is at $r_0=20M$. The top subplot presents the values of $\dot{E}_{\ell m}^{s,\infty}$, where the orange cross marks the results in this work, and the blue dot marks the results in \cite{Dyson:2025dlj}. At each $\ell$, $\dot{E}_{\ell m}^{s,\infty}$ decreases as $m$ increases, so higher the point higher the $m$. Since the modes with $\ell$ and $m$ of opposite parity and the $\ell=0,1$ modes do not contribute to the infinity flux, they are not displayed here. The bottom subplot shows the relative fractional differences in percentages between our results of $\dot{E}_{\ell m}^{s,\infty}$ and the corresponding ones in \cite{Dyson:2025dlj}.}
    \label{fig:mode_by_mode_flux}
\end{figure}

Besides the energy and angular momentum, there is an additional Noether charge $Q$ associated with the conserved current
\begin{equation}
    j_\mu^{\Phi}
    =-i\left(\bar{\Phi}\partial_\mu\Phi-\Phi\partial_\mu\bar{\Phi}\right)\,,
\end{equation}
such that
\begin{equation}
    Q=\int_{\Sigma} \sqrt{-g}j_{\Phi}^0\,d^3x\,,
\end{equation}
which determines the total number of particles within the cloud \cite{Annulli:2020lyc}. As shown in \cite{Brito:2023pyl, Dyson:2025dlj}, the mass $M_c$ and spin $S_c$ of the cloud are related to $Q$ by
\begin{equation} \label{eq:M_S_Q}
    M_c=\omega_cQ\,,\; S_c=m_cQ\,.
\end{equation}
The number flux of scalar particles at infinity $\dot{Q}^{\Phi,\infty}$ and the horizon $\dot{Q}^{\Phi,H}$ is given by \cite{Annulli:2020lyc, Brito:2023pyl, Dyson:2025dlj}
\begin{subequations}
\begin{align}
    & \dot{Q}^{\Phi,\infty}
    =-\lim _{r\rightarrow+\infty}r^2 
    \int j_r^{\Phi}\,d\Omega\,, \\
    & \dot{Q}^{\Phi,\mathrm{H}}
    =\lim_{r\rightarrow r_{+}}2Mr_{+}
    \int j_\mu^{\Phi}l^\mu\,d\Omega\,.
\end{align}
\end{subequations}
From Eq.~\eqref{eq:M_S_Q}, the flux in $Q$ will also induce an additional energy flux $\dot{M}_c=\omega_c\dot{Q}$ and angular momentum flux $\dot{S}_c=m_c\dot{Q}$ such that the total fluxes $\dot{E}^{s,\infty}$ and $\dot{L}^{s,\infty}$ due to scalar radiation are given by \cite{Dyson:2025dlj}
\begin{subequations}
\begin{align}
    & \dot{E}^{s,\infty}
    =\dot{E}^{\Phi,\infty}+\omega_c\dot{Q}^{\Phi,\infty}\,,\\
    & \dot{L}^{s,\infty}=
    \dot{L}^{\Phi,\infty}+m_c\dot{Q}^{\Phi,\infty}\,.
\end{align}
\end{subequations}

\begin{figure*}[t]
    \centering
    \includegraphics[width=\linewidth]{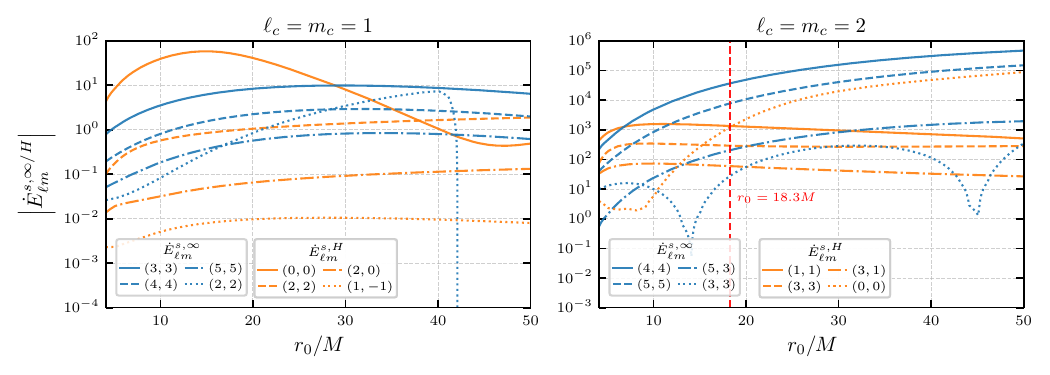}
    \caption{Leading mode-by-mode contributions to the infinity ($\dot{E}_{\ell m}^{s,\infty}$, blue lines) and horizon ($\dot{E}_{\ell m}^{s,H}$, orange lines) energy fluxes for the $(\ell_c=m_c=1,n_c=0)$ (left subplot) and $(\ell_c=m_c=2,n_c =0)$ (right subplot) clouds. Each line labeled with $(a,b)$ is for the $\ell=a,m=b$ mode of scalar radiation. Note that we have taken the absolute value of all the fluxes, where the infinity fluxes $\dot{E}_{\ell m}^{s,\infty}$ are always positive, while the horizon fluxes $\dot{E}_{\ell m}^{s,H}$ are negative for all the modes of the $\ell_c=m_c=1$ cloud in the plot and the $\ell=m=0,3$ modes of the $\ell_c=m_c=2$ cloud. The red vertical line marks the place ($r_0\approx18.3M$) inside which the horizon flux $\dot{E}_{\ell m}^{s,H}$ of the $\ell=m=1$ mode dominates over that of the $\ell=m=0$ mode for the $\ell_c=m_c=2$ cloud.} 
    \label{fig:flux_dominant_modes}
\end{figure*}

Using the decomposition of $\Phi^{(1,1)}$ in Eq.~\eqref{eq:phi_11_decomp} and their asymptotic behaviors determined by Eqs.~\eqref{eq:scalar_boundary} and \eqref{eq:scalar_sol_11}, one finds the mode-by-mode contribution to the fluxes to be \cite{Brito:2023pyl, Dyson:2025dlj}
\begin{subequations} \label{eq:fluxes_total_scalar}
\begin{align} 
    & \dot{E}_{\ell m}^{s,\infty} 
    =\lim_{r\rightarrow+\infty}2r^2m_g\omega_g\textrm{sgn}(\omega)
    \operatorname{Re}\left[\sqrt{\omega^2-\mu^2}\right]\left|
    {}_{0}\mathscr{R}_{\ell m\omega}^{\Phi}(r)\right|^2\,, 
    \label{eq:E_total_scalar_inf} \\
    & \dot{E}_{\ell m}^{s,\mathrm{H}} 
    =\lim_{r\rightarrow r_{+}}4Mr_{+}m_g\omega_g
    \left(\omega-m\omega_{+}\right)
    \left|{}_{0}\mathscr{R}_{\ell m\omega}^{\Phi}(r)\right|^2\,, 
    \label{eq:E_total_scalar_hor} \\
    & \dot{L}_{\ell m}^{s,\infty/H}
    =\Omega_{g}^{-1}\dot{E}_{\ell m}^{s,\infty/H}
    \label{eq:L_total_scalar} \,.
\end{align}
\end{subequations}
Since the angular momentum flux is completely determined by the energy flux in Eq.~\eqref{eq:L_total_scalar} for the circular orbits considered in this work, we will focus on the energy flux for the rest of this work. Inserting the ${}_{0}\mathscr{R}_{\ell m\omega}^{\Phi}(r)$ found in Sec.~\ref{sec:scalar_radiation} into Eq.~\eqref{eq:fluxes_total_scalar}, we obtain the horizon and infinity fluxes for a range of $r_0\in[4.1,50.1]M$ for the $\ell_c=m_c=1$ cloud. For a detailed comparison with the results in \cite{Dyson:2025dlj}, the infinity flux of each $(\ell,m)$ mode up to $\ell=10$ is shown in Fig.~\ref{fig:mode_by_mode_flux} for a secondary on a prograde orbit at $r_0=20M$ within the $\ell_c=m_c=1$ cloud. Note that Fig.~6 of \cite{Dyson:2025dlj} only pulls out a factor of $\epsilon^2 M_c/M$, while we have pulled out a factor of $\epsilon^2\zeta^2$ in Fig.~\ref{fig:mode_by_mode_flux}, so a factor of $(\mu M)^{6}$ needs to be removed from the former for comparison purposes. The results found via our approach agree with those obtained from the approach of \cite{Dyson:2025dlj}, with a relative fractional difference of $<5\%$ for all the modes present in Fig.~\ref{fig:mode_by_mode_flux}, except at $\ell=8,m=2$, where the relative fractional difference is $\sim 11.4\%$. This small discrepancy may arise from differences in constructing the radial source $\mathscr{S}_{\ell m}(r)$ in Eq.~\eqref{eq:scalar_sol_11} (our approach versus the approach in \cite{Dyson:2025dlj}), the precise boundary conditions and cutoff locations near the horizon and infinity used when calculating the homogeneous solutions ${}_{0}\mathscr{R}_{\ell m\omega}^{\textrm{up},\textrm{in}}(r)$, and the slightly different spin value used in \cite{Dyson:2025dlj} ($a=0.877M$ versus $a=0.88M$ here. When the flux becomes very small at $\ell=8,m=2$, higher precision is required for the source term and the homogeneous solutions when constructing Green's function to calculate the flux accurately, so the discrepancy between these two approaches becomes larger. However, since when $\ell=8,m=2$, the infinity flux is already about $10^{12}$ times smaller than the dominant contribution at $\ell=m=3$, this discrepancy is negligible. The mode-by-mode contribution to the infinity flux peaks at $\ell=m=3$ because the quadrupolar mode ($\ell_g=m_g=2$) of GWs driven by the secondary has the largest amplitude near spatial infinity. The same feature also appears for the $\ell_c=m_c=2$ cloud, where the infinity flux peaks at $\ell=m=4$, as shown in Fig.~\ref{fig:flux_dominant_modes}.

\begin{figure}[t]
    \centering
    \includegraphics[width=\linewidth]{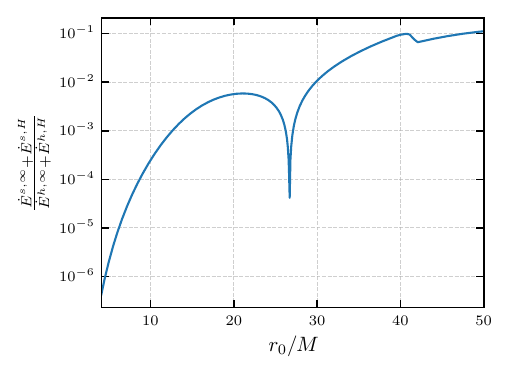}
    \caption{The ratio of the total energy flux due to scalar radiation $\dot{E}^{s,\infty}+\dot{E}^{s,H}$ to the total energy flux due to gravitational radiation $\dot{E}^{h,\infty}+\dot{E}^{h,H}$ for the $(\ell_c=m_c=1,n_c=0)$ scalar cloud of scalar mass $\mu M=0.3$ around a Kerr BH of spin $a=0.88M$. We have set the total mass of the cloud to be $M_c=10^{-4}M$. For the gravitational flux, we only consider the $\ell_g=m_g=2$ mode of the GW at $\mathcal{O}(\epsilon)$ driven by the secondary without incorporating the backreaction of the scalar cloud and scalar radiation. Notice this ratio is independent of the mass ratio $\epsilon$ to the leading order, as both the scalar and gravitational fluxes are $\propto\epsilon^2$.}
    \label{fig:flux_ratio}
\end{figure}

In Fig.~\ref{fig:total_flux_11}, we present the total energy fluxes at the horizon $\dot{E}_{\ell m}^{s,H}$ and at spatial infinity $\dot{E}_{\ell m}^{s,\infty}$ of the $\ell_c=m_c=1$ cloud, after summing up all the $(\ell,m)$ modes with $0\leq\ell\leq5$. Similar to \cite{Dyson:2025dlj}, we observe the sharp feature in the infinity flux near $r_0=41.66M$ due to the change in the near-infinity asymptotic behavior of $\Phi^{(1,1)}$ when $\omega^2-\mu^2$ changes sign during the secondary's inspiral. Nevertheless, as discussed in Sec.~\ref{sec:scalar_radiation}, this sharp feature is unlikely to be observed by detectors like LISA unless the scalar mass $\mu M$ is larger. Another feature worth noting is that at $r_0\lesssim 26.7M$, the horizon flux, which is always negative, starts to dominate over the infinity flux. Then, according to Eq.~\eqref{eq:energy_balance}, if the fluxes due to scalar radiation can balance those from gravitational radiation, the secondary could potentially ``float'' at this cross point \cite{Dyson:2025dlj, Zhang:2018kib}. However, as shown in Fig.~\ref{fig:flux_ratio}, the total energy flux due to scalar radiation is generally much weaker than the gravitational one until large orbital radii, so the secondary still migrates into the central supermassive BH, albeit more slowly. 

\begin{figure}[t]
    \centering
    \includegraphics[width=\linewidth]{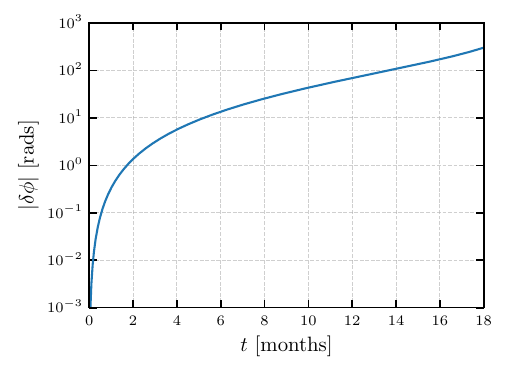}
    \caption{The absolute value of the dephasing $\delta\phi$ due to the scalar radiation of the $(\ell_c=m_c=1,n_c=0)$ scalar cloud as a function of time in months. The scalar field is of mass $\mu M=0.3$, and the cloud of mass $M_c=10^{-4}M$ is around a Kerr BH of spin $a=0.88M$ with $M=10^6M_{\odot}$. The mass ratio between the secondary and the Kerr BH is $\epsilon=10^{-5}$. At $t=0$, the secondary is at the orbital radius $r_0=10.6M$. Since the scalar radiation of the $\ell_c=m_c=1$ cloud slows down the secondary's inspiral when $r_0\lesssim 26.7M$, the dephasing $\delta\phi$ plotted here is negative.}
    \label{fig:dephasing}
\end{figure}

In Fig.~\ref{fig:total_flux_11}, we also compare our results to those obtained by \cite{Dyson:2025dlj}, where the infinity and horizon fluxes calculated from these two independent approaches have relative fractional differences $\lesssim6\%$ at most $r_0$. If we take the scalar field's mass to be $\mu M=0.3$ and the scalar cloud's total mass to be $M_c=10^{-4}M$, the total energy flux of the scalar radiation is $<10^{-3}$ times that of the gravitational radiation driven by the secondary when $r_0\le10M$ (see Fig.~\ref{fig:flux_ratio}). Thus, the discrepancies above will lead to a relative fractional error $\lesssim 10^{-5}$ in the total flux after including the gravitational flux due to the secondary. Although such an error looks tiny, it could potentially lead to a dephasing of $\mathcal{O}(1)$ rads after one year of observation \cite{Yunes:2009ef}, so further efforts are needed to reduce the discrepancy between these two approaches. Since considerable computational resources were spent in calculating the fluxes at $r_0>20M$ to reveal those robust features at large $r_0$, which are, however, outside the LISA band, it is more efficient to focus on $r_0\leq20M$ and improve the accuracy in that region. In Fig.~\ref{fig:dephasing}, we further show the dephasing due to the scalar radiation of this cloud when $M=10^6M_{\odot}$, $\mu M=0.3$, $M_c=10^{-4}M$, and $\epsilon=10^{-5}$ with a secondary starting the inspiral at $r_0=10.6M$. For the gravitational flux driven by the secondary, we only consider the dominant $\ell_g=m_g=2$ mode of the $\mathcal{O}(\epsilon)$ contribution to the GW without incorporating any effects of the scalar cloud, the latter of which will be investigated in follow-up works. Since this cloud slows down the secondary's inspiral when $r_0\lesssim26.7M$, the dephasing is always negative in Fig.~\ref{fig:dephasing}. After $18$ months of observation, the scalar radiation results in a dephasing of $\mathcal{O}(10^2)$ rads.

\begin{figure}[t]
    \centering
    \includegraphics[width=\linewidth]{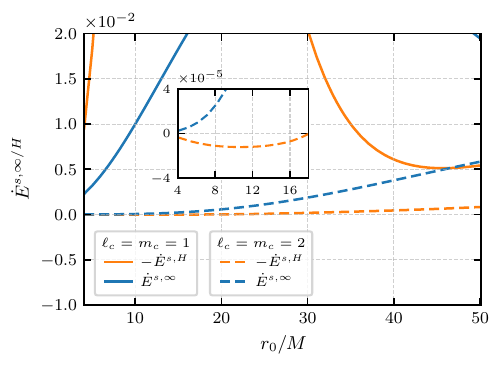}
    \caption{The infinity and horizon energy fluxes $\dot{E}^{s,\infty/H}$ associated with the scalar radiation $\Phi^{(1,1)}$ of the $(\ell_c=m_c=1,n_c=0)$ and $(\ell_c=m_c=2,n_c=0)$ scalar clouds around a Kerr BH of spin $a=0.88M$, where the scalar field is of mass $\mu M=0.3$. The blue lines stand for the infinity flux $\dot{E}^{s,\infty}$, while the orange lines are for the horizon flux $\dot{E}^{s,H}$, but with an overall minus sign since $\dot{E}^{s,H}$ is negative for most of $r_0$. The solid and dashed lines are the energy fluxes of the $\ell_c=m_c=1$ and $\ell_c=m_c=2$ clouds, respectively. The inset shows the range of $r_0$ inside which the horizon flux $\dot{E}^{s,H}$ of the $\ell_c=m_c=2$ cloud becomes positive.}
    \label{fig:total_flux_22}
\end{figure}

In addition, in Fig.~\ref{fig:total_flux_22}, we show the horizon and infinity energy fluxes due to the scalar radiation of the $\ell_c=m_c=2$ cloud, after summing up all the $(\ell,m)$ modes with $0\leq\ell\leq5$ and $-3\leq m\leq5$. Note that both the infinity and horizon fluxes of the $\ell_c=m_c=1$ cloud are much larger than those of the $\ell_c=m_c=2$ cloud, especially at small $r_0$. This feature is largely due to the fact that the background profile of the $\ell_c=m_c=2$ cloud has a broader profile with its peak at a larger radius than the $\ell_c=m_c=1$ cloud, as shown in Fig.~\ref{fig:scalar_profile}. Unlike the $\ell_c=m_c=1$ cloud, the total energy flux (i.e., $\dot{E}^{s,H}+\dot{E}^{s,\infty}$) of the $\ell_c=m_c=2$ cloud is always positive, which accelerates the inspiral of the secondary within such a cloud. Furthermore, inside $r_0\approx18.1M$, the central supermassive BH gains energy from the scalar radiation, resulting in positive horizon flux $\dot{E}^{s,H}$. This feature is mainly due to the $\ell=m=1$ mode of the scalar radiation. Similar to the $\ell_c=m_c=1$ cloud, the $\ell=m=0$ mode has the largest contribution to the horizon flux for most of the time, but the $\ell=m=1$ mode starts to dominate over the $\ell=m=0$ mode when $r_0\lesssim 18.3M$. This feature can be directly seen in Fig.~\ref{fig:flux_dominant_modes}, where we plot the first four leading modes contributing to the infinity and horizon fluxes for the two cloud configurations considered in this work. When $r_0\lesssim 18.1M$, the horizon flux of the $\ell=m=1$ mode is positive, as one can directly find from Eq.~\eqref{eq:E_total_scalar_hor} that $\dot{E}_{\ell m}^{s,H}>0$ when
\begin{equation}
    r_0<M^{1/3}\left(\frac{m_g}{m\omega_{+}-\omega_c}-a\right)^{2/3}\,.
\end{equation}
For the $\ell_c=m_c=2$ cloud, $\dot{E}_{\ell m}^{s,H}$ at $m=1$ remains positive as long as $r_0\lesssim 370M$, given a BH with spin $a=0.88M$, a scalar mass of $\mu M=0.3$, and the scalar cloud frequency $\omega_c$ in Eq.~\eqref{eq:omega_c_22}. In contrast, for the $\ell_c=m_c=1$ cloud, the maximum $r_0$ at which $\dot{E}_{\ell m}^{s,H}$ can be positive occurs when $m_g=-1$, but this threshold radius $r_0\approx 1.84M$ lies inside the horizon of a Kerr BH with spin $a=0.88M$, making the effect irrelevant. These differences in how various cloud configurations influence the secondary's evolution may help reveal the nature of scalar clouds around supermassive BHs via GW detections of EMRIs. 

\section{Discussion}
\label{sec:discussion}

In this work, we studied the evolution of an EMRI within an ultralight scalar cloud made up of a complex scalar field and formed via superradiance of the central supermassive BH. Built upon the MTF developed in \cite{Li:2022pcy, Hussain:2022ins}, we first extend this formalism, originally designed to study ringdown in beyond-GR theories \cite{Cano:2023tmv, Cano:2023jbk, Li:2023ulk, Wagle:2023fwl, Li:2025fci}, to the case of EMRIs within an ultralight scalar cloud. We then derive the modified Teukolsky equation and the sourced scalar field equation describing gravitational and scalar radiation, respectively, and lay down a detailed strategy for evaluating the source terms of both equations. We then focus on the scalar sector of the problem, leaving the complete calculation of the gravitational sector to follow-up work.

The source term in the scalar sector consists of terms from the background scalar cloud profile $\Phi^{(1,0)}$ and the perturbed metric $h_{\mu\nu}^{(0,1)}$, which is driven by the perturbations of the secondary body on the central BH. For the calculation of $\Phi^{(1,0)}$, we employ Leaver's method \cite{Leaver:1985ax, Dolan:2007mj}. For the comptuation of $h_{\mu\nu}^{(0,1)}$, we directly use the Lorenz-gauge reconstruction method and the associated {\texttt{Mathematica}} notebooks developed by \cite{Dolan:2021ijg, Dolan:2023enf, Wardell:2024yoi}. To simplify the source terms analytically, we project the source term to the NP basis and replace all the NP directional derivatives and spin coefficients with Chandrasekhar operators, which satisfy certain commutation relations. To convert the source term into an explicitly separable form, we further employ the method of \cite{Spiers:2024src} to decompose certain radial-angular factors into Fourier series of the polar angle. We then project this explicitly-separable source onto the spin-$0$ spheroidal harmonics to extract its radial part. Our semi-analytical treatment of the source term agrees very well with the results obtained through a more numerical approach \cite{Dyson:2025dlj}. Our procedure to compute the source term in the scalar sector can be directly applied to the gravitational sector. Due to the more complicated structure of the source term within the modified Teukolsky equation, this semi-analytical approach might prove more efficient and accurate when calculating the gravitational flux.

After obtaining the source term of the scalar radiation equation, we then used Green's function methods to solve for the radiation field. We calculated the scalar radiation emitted by two configurations of scalar clouds with scalar mass $\mu M=0.3$ around a Kerr BH of spin $a=0.88M$: the fundamental modes of the dipolar cloud ($\ell_c=m_c=1,n_c=0$) and the quadrupolar cloud ($\ell_c=m_c=2,n_c=0$). Consistent with the results of \cite{Dyson:2025dlj}, we observe a qualitative change in the scalar radiation of the dipolar cloud when the EMRI is located near $r_0\approx 41.66M$, which is driven by a shift in the asymptotic behavior of the $m=2$ modes as the difference of the square of the orbital frequency and the scalar mass, $\omega^2-\mu^2$, changes sign. For the quadrupolar cloud, the transition occurs at a larger EMRI orbital separation $r_0$ and produces significantly weaker scalar radiation, assuming equal masses with the dipole cloud.

We use the scalar radiation field $\Phi^{(1,1)}$ to further compute the energy fluxes at spatial infinity and at the BH horizon, carried away by the scalar radiation. Our results for the dipole cloud are highly consistent with those in \cite{Dyson:2025dlj}, including a sharp decrease of the infinity flux at $r_0\approx 41.66M$ as $r_0$ increases. Consistent with \cite{Dyson:2025dlj}, we also find a threshold EMRI orbital radius $r_0\approx26.7M$, within which the negative scalar horizon flux dominates over the scalar infinity flux, yielding a net negative energy flux due to scalar radiation. In this case, the scalar radiation slows down the inspiral of the secondary within $r_0\approx26.7M$. In contrast, the net energy flux of the scalar radiation generated by the quadrupolar cloud is always positive, accelerating the inspiral of the secondary. In addition, unlike the dipole cloud, when $r_0\lesssim18.1M$, the scalar horizon flux becomes positive, so the scalar radiation deposits energy into the central supermassive BH. These features in how different configurations of scalar clouds affect the secondary may provide us with a novel way to probe the structure of ultralight scalar clouds with future GW observations. 

We have here presented a fully relativistic analysis of scalar radiation and its associated energy fluxes sourced by EMRIs embedded in ultralight scalar clouds, taking the first steps toward a comprehensive modeling of EMRIs in such environments. In our follow-up work, we will apply this framework, which is based on the MTF \cite{Li:2022pcy, Hussain:2022ins, Cano:2023tmv}, to study gravitational radiation and explore its implications for observational probes of ultralight scalar clouds. While we have focused on two simple cloud configurations at fixed BH spin and scalar mass, our methods can be readily applied to a broader parameter space to study how different cloud profiles interact with EMRIs, as we have partially explored in this work. Although this work, along with its follow-up, aims at modeling EMRIs within ultralight scalar clouds, the procedures and techniques we developed are general and can be adapted to EMRIs in other astrophysical environments or beyond-GR theories.

\section{Acknowledgements} 
\label{sec:acknowledgements}
\appendix

We are grateful to Conor Dyson and Thomas F.M. Spieksma for sharing the data and assisting with the verification of calculations in \cite{Dyson:2025dlj}. We thank Richard Brito and Yanbei Chen for helpful discussions related to this project. We also thank Sam Dolan for providing the Kerr-Lorenz-Circ {\texttt{Mathematica}} code developed in \cite{Dolan:2021ijg, Dolan:2023enf, Wardell:2024yoi} and clarifying its usage. We appreciate the insightful comments on the final draft of this manuscript from Conor Dyson, Thomas F.M. Spieksma, and Richard Brito. D.~L. and N.~Y. acknowledge support from the Simons Foundation through Award No. 896696, NSF Grant No. PHY-2207650, and the National Aeronautics and Space Administration through award 80NSSC22K0806. P.~B. acknowledges support from the Dutch Research Council (NWO) with file number OCENW.M.21.119. C.~W.'s  research is supported by the Simons Foundation (Award No. 568762), the Brinson Foundation, and the National Science Foundation (via Grants No. PHY-2011961 and No. PHY-2011968). This work makes use of the Kerr-Lorenz-Circ {\texttt{Mathematica}} code developed by Sam Dolan \cite{Dolan:2021ijg, Dolan:2023enf, Wardell:2024yoi} and the Black Hole Perturbation Toolkit \cite{BHPToolkit}. Our calculations use the Illinois Campus Cluster, a computing resource that is operated by the Illinois Campus Cluster Program (ICCP) in conjunction with the National Center for Supercomputing Applications (NCSA), and is supported by funds from the University of Illinois Urbana-Champaign (UIUC). Some calculations were also conducted in the Resnick High Performance Computing Center, a facility supported by the Resnick Sustainability Institute at the California Institute of Technology.

\section{Chandrasekhar operators} 
\label{appendix:chandrasekhar_operators}

In this appendix, we list the commutation relations between the Chandrasekhar operators $\left\{\mathcal{D}_{n},\mathcal{D}^{\dagger}_{n},
\mathcal{L}_{n},\mathcal{L}^{\dagger}_{n}\right\}$ in Eq.~\eqref{eq:chandra_ops} and the functions $\left\{\Delta(r),\sin{\theta},\Gamma(r,\theta),\bar{\Gamma}(r,\theta)\right\}$. These commutation relations have already been studied at many places, such as \cite{Chandrasekhar_1983, Ma:2024qcv}, but we provide them here for completeness, i.e.,
\begin{subequations} \label{eq:commutation_chandra}
\begin{align}
    \mathcal{D}_{n}\left(\Delta^mf\right)
    =& \;\Delta^{m}\mathcal{D}_{n+m}f\,, \\
    \mathcal{D}_{n}\left(\sin^m{\theta}f\right)
    =& \;\sin^{m}{\theta}\mathcal{D}_{n}f\,, \\
    \mathcal{D}_{n}\left(\Gamma^m f\right)
    =& \;\Gamma^{m}\mathcal{D}_{n}f+m\Gamma^{m-1}f\,, \\
    \mathcal{D}_{n}\left(\bar{\Gamma}^mf\right)
    =&\;\bar{\Gamma}^{m}\mathcal{D}_{n}f+m\bar{\Gamma}^{m-1}f\,, \\
    \mathcal{L}_{n}\left(\Delta^mf\right)
    =& \;\Delta^{m}\mathcal{L}_{n}f\,, \\
    \mathcal{L}_{n}\left(\sin^m{\theta}f\right)
    =& \;\sin^{m}{\theta}\mathcal{L}_{n+m}f\,, \\
    \mathcal{L}_{n}\left(\Gamma^mf\right)
    =& \;\Gamma^{m}\mathcal{L}_{n}f
    -ima\Gamma^{n-1}\sin{\theta}f\,, \\
    \mathcal{L}_{n}\left(\bar{\Gamma}^mf\right)
    =&\;\bar{\Gamma}^{m}\mathcal{L}_{n}f 
    +ima\Gamma^{n-1}\sin{\theta}f\,,
\end{align}
\end{subequations}
where $\Delta\equiv\Delta(r)=r^2-2Mr+a^2$ is purely radial, and $\Gamma\equiv\Gamma(r,\theta)=r+ia\cos{\theta}$. The relations between $\left\{\mathcal{D}^{\dagger}_{n},\mathcal{L}^{\dagger}_{n}\right\}$ and those functions can be obtained by simply replacing $\left\{\mathcal{D}_{n},\mathcal{L}_{n}\right\}$ with $\left\{\mathcal{D}^{\dagger}_{n},\mathcal{L}^{\dagger}_{n}\right\}$ in Eq.~\eqref{eq:commutation_chandra}. Since the operators $\left\{\mathcal{D}_{n},\mathcal{D}^{\dagger}_{n}\right\}$ are purely radial, and the operators $\left\{\mathcal{L}_{n},\mathcal{L}^{\dagger}_{n}\right\}$ are purely angular, the former commutes with the latter.

Furthermore, the operators in Eq.~\eqref{eq:operators_shorthand} on the Kerr background can be directly mapped to the Chandrasekhar operators, i.e.,
\begin{align} 
    D_{[p,q,u,v]}
    &=\mathcal{D}_0-\left(\frac{v}{\Gamma}
    +\frac{u}{\bar{\Gamma}}\right)\,, \nonumber\\
    \boldsymbol{\Delta}_{[p,q,u,v]}
    &=-\frac{\Delta}{\Gamma\bar{\Gamma}}
    \left(\mathcal{D}^{\dagger}_{-(u+v)/2}
    +\frac{q+v}{\Gamma}+\frac{p+u}{\bar{\Gamma}}\right)\,, 
    \nonumber\\
    \delta_{[p,q,u,v]}
    &=\frac{1}{\sqrt{2}\Gamma}
    \left[\mathcal{L}^{\dagger}_{(q-p)/2}-ia\sin{\theta}
    \left(\frac{p+u}{\Gamma}
    +\frac{v}{\bar{\Gamma}}\right)\right]\,, \nonumber\\
    \bar{\delta}_{[a,b,c,d]}
    &=\frac{1}{\sqrt{2}\bar{\Gamma}}
    \left[\mathcal{L}_{(q-p)/2}+ia\sin{\theta}
    \left(\frac{p+u}{\Gamma}
    +\frac{v}{\bar{\Gamma}}\right)\right]\,.
\end{align}   
These operators can be further mapped to the Geroch-Held-Penrose (GHP) operators \cite{Geroch:1973am}, as shown in \cite{Pound:2021qin, Ma:2024qcv}.

\section{Complete form of the source term $\mathcal{S}_{\Phi}^{(1,1)}$} 
\label{appendix:source_terms}

In this appendix, we provide the complete form of the source term $\mathcal{S}_{\Phi}^{(1,1)}$ in Eq.~\eqref{eq:EOM_scalar_11_source_Lorentz} in terms of the Chandrasekhar operators $\left\{\mathcal{D}_{n},\mathcal{D}^{\dagger}_{n},
\mathcal{L}_{n},\mathcal{L}^{\dagger}_{n}\right\}$ in Eq.~\eqref{eq:chandra_ops} and the functions $\left\{\Delta(r),\sin{\theta},\Gamma(r,\theta),\bar{\Gamma}(r,\theta)\right\}$. We have found that the $(\ell,m)$ component of $\mathcal{S}_{\Phi}^{(1,1)}$ is
\begin{align} \label{eq:source_complete}
    \Gamma\bar{\Gamma}\mathcal{S}_{\Phi}^{\ell m(1,1)}
    =& \;e^{-i\omega t+im\phi}\left({}_{2}\hat{\mathcal{S}}_{\Phi}^{(1,1)}
    +{}_{1}\hat{\mathcal{S}}_{\Phi}^{(1,1)}
    +{}_{0}\hat{\mathcal{S}}_{\Phi}^{(1,1)}\right) \nonumber\\
    & \;{}_{0}R^{\Phi}_{\ell_cm_c\omega_c}(r)
    {}_{0}S^{\Phi}_{\ell_c m_c\omega_c}(\theta)\,,
\end{align}
where the factor $\Gamma\bar{\Gamma}$ is to make the scalar field equation in the NP basis explicitly separable, and ${}_{s}\hat{\mathcal{S}}_{\Phi}^{(1,1)}$ are operators dependent on the spin-$s$ parts of the reconstructed metric $h_{\tilde{a}\tilde{b}}^{(0,1)}$, i.e.,
\begin{widetext}
\begin{subequations}
\begin{align} \label{eq:source_components}
    {}_{2}\hat{\mathcal{S}}_{\Phi}^{(1,1)}
    =& \;\frac{1}{4\Gamma\bar{\Gamma}}
    \left({}_{+2}R^{m_{+}m_{+}}{}_{+2}Y\mathcal{L}_{-1}\mathcal{L}_0
    +{}_{-2}R^{m_{-}m_{-}}{}_{-2}Y
    \mathcal{L}^{\dagger}_{-1}\mathcal{L}^{\dagger}_0
    \right)\,, \\
    {}_{1}\hat{\mathcal{S}}_{\Phi}^{(1,1)}
    =& \;\frac{\Delta}{2\Gamma^3\bar{\Gamma}}
    \left[{}_{+1}R^{l_{+}m_{+}}{}_{+1}Y
    \left(ia\sin{\theta}\mathcal{D}_{0}^{\dagger}
    -\mathcal{L}_{0}\right)
    +{}_{-1}R^{l_{-}m_{-}}{}_{-1}Y
    \left(ia\sin{\theta}\mathcal{D}_{0}
    -\mathcal{L}_{0}^{\dagger}\right)\right] \nonumber\\
    & \;+\frac{\Delta}{2\Gamma^2\bar{\Gamma}}
    \left({}_{+1}R^{l_{+}m_{+}}{}_{+1}Y\mathcal{L}_{0}\mathcal{D}_0^{\dagger}
    +{}_{-1}R^{l_{-}m_{-}}{}_{-1}Y\mathcal{L}_{0}^{\dagger}\mathcal{D}_0\right)
    +\frac{\Delta}{2\Gamma\bar{\Gamma}^2}
    \left({}_{-1}R^{l_{+}m_{-}}{}_{-1}Y
    \mathcal{L}_{0}^{\dagger}\mathcal{D}_0^{\dagger}
    +{}_{+1}R^{l_{-}m_{+}}{}_{+1}Y\mathcal{L}_{0}\mathcal{D}_0\right) \nonumber\\
    & \;-\frac{\Delta}{2\Gamma\bar{\Gamma}^3}
    \left[{}_{-1}R^{l_{+}m_{-}}{}_{-1}Y
    \left(ia\sin{\theta}\mathcal{D}_{0}^{\dagger}
    +\mathcal{L}_{0}^{\dagger}\right)
    +{}_{+1}R^{l_{-}m_{+}}{}_{+1}Y\left(ia\sin{\theta}\mathcal{D}_{0}
    +\mathcal{L}_{0}\right)\right]\,, \\
    {}_{0}\hat{\mathcal{S}}_{\Phi}^{(1,1)}
    =& \;\frac{1}{4\Gamma^3\bar{\Gamma}^2}
    \left(\Gamma^2\bar{\Gamma}^2{}_{0}R^{h}
    -2\,{}_{0}R^{l_{+}l_{-}}\right){}_{0}Y
    \left[\Delta\left(\mathcal{D}_{0}
    +\mathcal{D}_{0}^{\dagger}\right)
    +ia\sin{\theta}\left(\mathcal{L}_0
    +\mathcal{L}_0^{\dagger}\right)\right] \nonumber\\
    & \;+\frac{1}{4\Gamma^2\bar{\Gamma}^3}
    \left(\Gamma^2\bar{\Gamma}^2{}_{0}R^{h}
    -2\,{}_{0}R^{l_{+}l_{-}}\right){}_{0}Y
    \left[\Delta\left(\mathcal{D}_{0}
    +\mathcal{D}_{0}^{\dagger}\right)
    -ia\sin{\theta}\left(\mathcal{L}_0
    +\mathcal{L}_0^{\dagger}\right)\right] \nonumber\\
    & \;+\frac{1}{4\Gamma^2\bar{\Gamma}^2}
    \left(\Gamma^2\bar{\Gamma}^2{}_{0}R^{h}
    -{}_{0}R^{l_{+}l_{-}}\right){}_{0}Y
    \left(\mathcal{L}_1\mathcal{L}_0^{\dagger}
    +\mathcal{L}_1^{\dagger}\mathcal{L}_0\right)
    +\frac{\Delta^2}{4\Gamma\bar{\Gamma}}{}_{0}Y
    \left({}_{0}R^{l_{-}l_{-}}\mathcal{D}_{0}\mathcal{D}_{0}
    +{}_{0}R^{l_{+}l_{+}}\mathcal{D}_{0}^{\dagger}
    \mathcal{D}_{0}^{\dagger}\right)\,,
\end{align}
\end{subequations}  
\end{widetext}
where we have dropped the arguments and the subscripts labeling the modes of all the functions for simplicity. Recall that ${}_{s}R^{\tilde{a}\tilde{b}}\equiv{}_{s}R^{\tilde{a}\tilde{b}}_{\ell_gm_g\omega_g}(r)$ and ${}_{s}Y\equiv{}_{s}Y_{\ell_gm_g}(\theta)$ are the radial and angular parts of the reconstructed metric $h_{\tilde{a}\tilde{b}}^{(0,1)}$ defined in Eq.~\eqref{eq:h_decomp}, respectively, where the indices $\tilde{a},\tilde{b}$ label the components in the unnormalized Kinnersley tetrad $\tilde{e}_{a}^{\mu}=\{l_{+}^{\mu},l_{-}^{\mu},m_{+}^{\mu},m_{-}^{\mu}\}$ defined in Eq.~\eqref{eq:tetrad}. The function ${}_{0}R^{h}\equiv{}_{0}R^{h}_{\ell_gm_g\omega_g}(r)$ is the radial part of the trace $h$ of the reconstructed metric $h_{\tilde{a}\tilde{b}}^{(0,1)}$.

\bibliographystyle{apsrev4-2}
\bibliography{reference}

\end{document}